\documentclass[aps,twocolumn,prb,showpacs,superscriptaddress,floatfix]{revtex4}
\usepackage{graphicx,epsfig}
\usepackage{newlfont,bm,dcolumn,longtable}
\usepackage{amssymb,amsfonts,amsmath}
\usepackage{color}

\newcommand{\MT}{\mathsf M}
\newcommand{\MTs}{\mathsf M_\mathrm{s}}
\newcommand{\Ms}{\vec M_\mathrm{s}}
\newcommand{\M}{\vec M}
\newcommand{\muT}{\mu^\mathrm{T}}
\newcommand{\muTs}{\mu^\mathrm{T}_\mathrm{s}}
\newcommand{\muV}{\mu^\mathrm{V}}
\newcommand{\muVs}{\mu^\mathrm{V}_\mathrm{s}}
\newcommand{\chiT}{\chi^\mathrm{T}}
\newcommand{\chiTs}{\chi^\mathrm{T}_\mathrm{s}}
\newcommand{\chiV}{\chi^\mathrm{V}}
\newcommand{\chiVs}{\chi^\mathrm{V}_\mathrm{s}}
\newcommand{\vphi}{\vec\phi}
\newcommand{\even}{\mathrm{even}}
\newcommand{\odd}{\mathrm{odd}}
\newcommand{\ii}{{\mathrm i}}
\newcommand{\ee}{{\mathrm e}}
\newcommand{\oo}{{\mathrm o}}
\newcommand{\dd}{{\mathrm d}}
\newcommand{\st}{{\mathrm s}}

\begin{document}

\title{Phase diagram of the (bosonic) Double-Exchange Model}

\author{J.~L.~Alonso}
\affiliation{Departamento de F\'{\i}sica Te\'orica,
Facultad de Ciencias, \\
Universidad de Zaragoza, 50009 Zaragoza, SPAIN}
\affiliation{Instituto de Biocomputaci\'on y  F\'{\i}sica de
Sistemas Complejos (BIFI)}
\author{A.~Cruz}
\affiliation{Departamento de F\'{\i}sica Te\'orica,
Facultad de Ciencias, \\
Universidad de Zaragoza, 50009 Zaragoza, SPAIN}
\affiliation{Instituto de Biocomputaci\'on y  F\'{\i}sica de
Sistemas Complejos (BIFI)}
\author{L.~A.~Fern\'andez}
\affiliation{Departamento de F\'{\i}sica Te\'orica, 
Facultad de Ciencias F\'{\i}sicas,\\
Universidad Complutense, 28040 Madrid, SPAIN}
\affiliation{Instituto de Biocomputaci\'on y  F\'{\i}sica de
Sistemas Complejos (BIFI)}
\author{S.~Jim\'enez}
\affiliation{Departamento de F\'{\i}sica Te\'orica,
Facultad de Ciencias, \\
Universidad de Zaragoza, 50009 Zaragoza, SPAIN}
\affiliation{Instituto de Biocomputaci\'on y  F\'{\i}sica de
Sistemas Complejos (BIFI)}
\author{V.~Mart\'\i{}n-Mayor} 
\affiliation{Departamento de F\'{\i}sica Te\'orica, 
Facultad de Ciencias F\'{\i}sicas,\\
Universidad Complutense, 28040 Madrid, SPAIN}
\affiliation{Instituto de Biocomputaci\'on y  F\'{\i}sica de
Sistemas Complejos (BIFI)}
\author{J.~J.~Ruiz-Lorenzo}
\affiliation{Departamento de F\'{\i}sica, Facultad de Ciencias,\\
Universidad de Extremadura, 06071 Badajoz, SPAIN}
\affiliation{Instituto de Biocomputaci\'on y  F\'{\i}sica de
Sistemas Complejos (BIFI)}
\author{A.~Taranc\'on}
\affiliation{Departamento de F\'{\i}sica Te\'orica,
Facultad de Ciencias, \\
Universidad de Zaragoza, 50009 Zaragoza, SPAIN}
\affiliation{Instituto de Biocomputaci\'on y  F\'{\i}sica de
Sistemas Complejos (BIFI)}
\date{\today}

\begin{abstract}
The phase diagram of the simplest approximation to Double-Exchange
systems, the bosonic Double-Exchange model with antiferromagnetic
super-exchange coupling, is fully worked out by means of Monte Carlo
simulations, large-$N$ expansions and Variational Mean-Field
calculations. We find a rich phase diagram, with no first-order phase
transitions. The most surprising finding is the existence of a segment
like ordered phase at low temperature for intermediate AFM coupling
which {\em cannot be detected in neutron-scattering experiments}.
This is signaled by a maximum (a cusp) in the specific
heat. Below the phase-transition, only short-range ordering would be
found in neutron-scattering.  Researchers looking for a Quantum
Critical Point in manganites should be wary of this
possibility. Finite-Size Scaling estimates of critical exponents are
presented, although large scaling corrections are present in the
reachable lattice sizes.
\end{abstract}
\maketitle

\section{Introduction}

There are at least two motivations for studying the spin-only version
of the Double-Exchange (DE) models. On the one hand one, has its
relationship with the Colossal Magnetorresistance effect
(CMR).\cite{CVM99,DAGOTTO,DAGOTTOBOOK} On the other hand, in this problem some
puzzles arise\cite{RAIZLETTER} with the Universality
Hypothesis,\cite{AMIT} that deserve a detailed study. Let us start
addressing the first aspect.

CMR has renewed the interest in Double-Exchange (DE)
systems.\cite{DEM} The typical CMR manganites are
$\mathrm{La}_{1-x}AE_x\mathrm{Mn}_{1-y}\mathrm{O}_3\, , $ where {\it
AE}\,=\,Ca,\,Sr in the range $0.2<x<0.5$. It is believed that the
relevant degrees of freedom\cite{CVM99} are the localized $S=3/2$
Mn$^{3+}$ core spins, and the $\mathbf{e_g}$ holes. The Mn$^{3+}$ ions
form a single cubic lattice and, besides the DE mechanism, interact
through an antiferromagnetic (AFM) superexchange coupling. The
relatively high spin of the Mn$^{3+}$ core suggests to treat them as
classical spins, $\vec \phi_i$.  Although phonons are believed to be
crucial for the CMR effect,\cite{Verges01} manganites display a very
rich magnetic phase diagram which can be addressed neglecting lattice
effects.\cite{Alonso01} In spite of these simplifications, and of the
introduction of powerful new
tools,\cite{Alonso00a,Alonso00b,Alonso00c} the numerical study of the
DE model in large lattices beyond Mean Field Approximation is out of
reach for present day computers. Yet, finite-size effects in these
systems are unusually large.\cite{Alonso01} The need to obtain
reliable predictions has made people to further simplify models,
replacing $\mathbf{e_g}$ holes by an effective interaction among the
localized $S=3/2$ Mn$^{3+}$ core spins. Indeed, a simple
calculation\cite{DEM} shows that the kinetic energy of the electrons
depends on the relative orientation of neighboring Mn$^{3+}$ core
spins as $\sqrt{1+\vec \phi_i \vec \phi_j}$. This substitution of a
simpler spin-only problem in place of the very difficult electronic
problem lies at the heart of several theoretical analyses (see e.g. de
Gennes in Ref.~\onlinecite{DEM}) and numerical simulations.\cite{KuMa}
In spite of this, up to our knowledge there is only one detailed
previous study\cite{TSAILANDAU} of the phase-diagram of the bosonic DE
model. That study predicted the existence of a disordered (PM) phase
at very low temperatures for intermediate super-exchange
coupling. This is very remindful of the presence of a Quantum-Critical
Point\cite{QCP} which is believed to be of importance for the CMR
phenomenon,\cite{Burgy01} and has been predicted to occur in
manganites by some model calculations.\cite{Alonso02} The experimental
characterization of this Quantum Critical Point is a wedge of
paramagnetic (PM) phase, maybe glassy,\cite{DAGOTTO03} that at zero
temperature becomes a single point separating two ordered
phases.\cite{Burgy01} The glassy wedge would be created by
disorder,\cite{Burgy01} and would be separated from the Paramagnetic
state at high-temperature scale $T^*$. Maybe the most surprising
result of the here presented analysis is that this glassy wedge could
not be PM or glassy at all, but ordered in a segment-like
way\cite{RP2D3,ROMANO,SHROCK} (as in liquid crystals). This ordering
will be referred to in the following as RP$^2$ (real projective
space).  As we shall show, the RP$^2$ phase cannot be
detected in neutron-scattering experiments (although a short-range
ordering will be present). Nevertheless, the phase-transition can be
studied experimentally using the specific-heat, that should present a
maximum (furthermore, a cusp) at the critical temperature. Indeed, the
thermal critical exponent is predicted\cite{RAIZLETTER} to be
$\nu=0.78(2)$ which implies $\alpha=-0.34$, and hence the cusp
behavior follows. Another bonus of our simplified model is that it
allows us to study qualitatively (see subsection~\ref{HSUBSECT}) the
funny interplay between ferromagnetism, antiferromagnetism,
temperature and applied magnetic field in low-doped
La$_{1-x}$Sr$_x$MnO$_3$.\cite{Uhlen,Nojiri,Wagner,Shun}

Let us now address Universality.  A common wisdom is that the critical
properties of a system are given by its dimensionality and the local
properties (i.e. near the identity element) of the coset space $\cal
G/\cal H$, where $\cal G$ is the symmetry group of the Hamiltonian
(the symmetry of the high temperature phase) and $\cal H$ is the
remaining symmetry group of the broken phase (low temperature). So,
systems with locally isomorphic $\cal G/\cal H$ belong to the same
Universality Class.  This seems to be true in perturbation theory,
where the observables are computed by doing series expansions around
the identity element of $\cal G/\cal H$. In this picture, a phase
transition of a vector model, with O(3) global symmetry and with an
O(2) low temperature phase symmetry, in three dimensions must belong
to the O(3)/O(2) scheme of symmetry breaking (classical Heisenberg
model). In addition, if $\cal H$=O(1)=Z$_2$ is the remaining symmetry,
the corresponding scheme should be\cite{AZARIA} O(4)/O(3) which is locally
isomorphic to O(3)/O(1).

Hence, it is interesting to check if the global properties of the
coset space $\cal G/\cal H$ are relevant or not to the phase
transition.  The common wisdom has been challenged in the past by the
so-called chiral models.\cite{KAWAMURA} However, the situation is
still hotly debated: some authors believe that the chiral transitions
are weakly first-order,\cite{NOCHIRAL} while others
claim\cite{ChiralN23} that the Chiral Universality Class exists,
implying the relevance of the global properties of $\cal G/\cal H$.
On the other hand, we do not have any doubt about the second order
nature of the PM-RP$^2$ transition. A detailed study of the critical
exponents was recently published\cite{RAIZLETTER} in Letter form. In
the present work, we perform a detailed Monte Carlo, Mean Field and
large-$N$ study of the phase diagram. A thorough study is performed
of the RP$^2$ phase. We shall confirm that the pattern of symmetry
breaking is O(3)/O(2), implying a violation of Universality for which
we do not have any plausible explanation.

The layout of the rest of this paper is as follows. In
section~\ref{MODELSECT} we define the model, study the phase diagram
at zero temperature, and define the order parameters and observables
measured in the Monte Carlo simulations. The Mean-Field calculation is
explained in section~\ref{MEANFIELDSECT}, where we also report the
results of the Large-$N$ analysis. In section~\ref{MONTECARLOSECT} we
present our Monte Carlo results. We start determining the global phase 
diagram in subsection~\ref{PHDI}.
The RP$^2$ phase is investigated in more detail in
subsection~\ref{RP2SUBSECT}, while the effects of a magnetic field on
conductivity close to a ferromagnetic-antiferromagnetic transition is
considered in subsection~\ref{HSUBSECT}. We present our conclusions in
section~\ref{CONCLUSIONSSECT}. We complement the paper with two
appendices.  Appendix~\ref{LARGEN} contains the details about the
large-$N$ calculation.  In Appendix~\ref{FOURTH} the reader will find
the Mean-Field phase diagram as obtained from the fourth-order
expansion of the Free-Energy.

\section{The model}\label{MODELSECT}

\subsection{The Hamiltonian}

We define a system of spins $\{\vphi_i\}$ living in a three
dimensional cubic lattice of size $L$ (and volume $V=L^3$) with
periodic boundary conditions. The spins are three-component real unit
vectors. We consider the Hamiltonian
\begin{equation}
H=-\sum_{<i,j>}\left(J\vphi_i\cdot\vphi_j+\sqrt{1+\vphi_i\cdot\vphi_j}\right)
\ ,
\label{accion}
\end{equation}
where the sum is extended to all pairs of nearest neighbors and we
consider only $J<0$.  Notice that we will measure temperature in units
of the double exchange constant. The cubic lattice is bipartite,
therefore we shall call the lattice site $i$ {\em even} or {\em odd}
according to the parity of the sum of its coordinates, $x_i+y_i+z_i$\,.

We will consider the system at a temperature $T$, the partition
function being
\begin{equation}
Z=\int\prod_i \dd \vphi\ \ee^{-H/T}\ , 
\end{equation}
where the integration measure is the standard measure on the unit sphere.

\subsection{Phase diagram at zero temperature}

As usual, the study of the phase diagram begins with an understanding
of the ordered phases at zero temperature.  We can write in a compact
way our original Hamiltonian:
\begin{equation}
H=-\sum_{<i,j>} {\cal
V}(\vphi_{i}\cdot \vphi_{j})\;,
\label{zero}
\end{equation}
where
\begin{equation}
{\cal V}(y)=J y+\sqrt{1+y} \,,
\end{equation}
and clearly $y\in[-1,1]$.  In the limit of zero temperature, the only
configurations which contribute to the partition function are those
which provide a maximum of ${\cal V}(y)$. If, as confirmed by the
Monte Carlo simulations, the spin texture itself is bipartite, the
value of $y$ will be uniform through the lattice. Thus, a simple
computation yields that the maxima of ${\cal V}(y)$ are at the
following values of $y$ (denoted by $y_\mathrm{max}$)
\begin{equation}
\setlength{\extrarowheight}{4pt}
y_\mathrm{max}=\left\{
\begin{array}{c c c}
1 & \mathrm{for} & J\ge -\frac{1}{2 \sqrt{2}} \\
-1+\frac{1}{4 J^2}& \mathrm{for} & J\le -\frac{1}{2 \sqrt{2}} .
\end{array}
\right.
\end{equation}
It is clear that $y_\mathrm{max}=1$ corresponds to a ferromagnetic
state and that in the $J \to -\infty$ limit we reach an
anti-ferromagnetic one ($y_\mathrm{max}=-1$). The intermediate values
of $y_\mathrm{max}$ correspond to a ferrimagnet if $0<y_\mathrm{max}
<1$ and to an anti-ferrimagnet when $-1<y_\mathrm{max}< 0$. The
physical picture is as follows: the spins in the, say, even sublattice
are all parallel along (for instance) the Z-axis. On the other hand
the odd spins lie on a cone forming an angle $\theta$
($\cos\theta=y_\mathrm{max}$) with the Z-axis.

The corresponding free energy is just
\begin{equation}
\setlength{\extrarowheight}{4pt}
f(J)=\left\{
\begin{array}{c c c}
\sqrt{2}+J & \mathrm{for} & J\ge -\frac{1}{2 \sqrt{2}}\,,\\
-\frac{1}{4 J}-J & \mathrm{for} & J\le -\frac{1}{2 \sqrt{2}}\,.
\end{array}
\right.
\end{equation}

Hence we have the following phase transitions:
\begin{enumerate}

\item \underline{Ferromagnetic-ferrimagnetic} at $J=-1/\sqrt{8}$. 
  It is easy to check that $\dd f/\dd J$ is continuous
  at \hbox{$J=-1/\sqrt{8}$} but $\dd^2f/\dd J^2$ is
  discontinuous. Hence, according to the standard Erhenfest
  classification, we have a second order phase transition.

\item \underline{Ferrimagnetic-antiferrimagnetic} at $J=-1/2$,
 where the free-energy is $C^\infty$. At this special point
 $y_\mathrm{max}$ changes from positive to negative. The fact that
 $y_\mathrm{max}=0$ implies that one can reverse every single spin
 independently of the others without changing the energy (more
 pedantically, one finds a dynamically generated Z$_2$ gauge
 symmetry\cite{RAIZLETTER}).

\item The limiting value $y_\mathrm{max}=-1$ that corresponds to an
  \underline{antiferromagnet} rather than an antiferrimagnet is
  reached only at $J=-\infty$.

\end{enumerate}
The transition 2 (Ferrimagnet-antiferrimagnet) needs further
discussion. We can expand the Hamiltonian around the minimum $y=0$,
and we obtain
\begin{equation}
\cal{V}(y) = {1}-\frac{1}{8} y^2+{\cal O}(y^3)\,.
\end{equation}
Thus, close to $J=-0.5$ and $T=0$ one has, neglecting constant terms, 
\begin{equation}
H= \frac{1}{8}\sum_{<i,j>} (\vphi_{i}\cdot \vphi_{j})^2 + 
{\cal O}\left((\vphi_{i}\cdot \vphi_{j})^3\right)\,,
\end{equation}
which corresponds to an {\em anti}-ferromagnetic RP$^2$-theory. The
minimum energy configuration satisfies $y=0$.  Hence, we obtain that
the ferrimagnet-antiferrimagnet transition occurs at zero
temperature via a RP$^2$ state at a single point. We shall see that at
finite temperature the RP$^2$ phase occupies a region (rather than a
single point) of the phase diagram.
 
From the previous analysis at zero temperature, one expects to find
the following phases at finite temperature:

\begin{itemize}
\item{PM:}
the usual disordered state, where all the symmetries of the
model are preserved.
\item{FM:} a standard ferromagnetic ordering, i.e., the spin
fluctuates around $(0,0,1)$.
\item{AFM:} a standard anti-ferromagnetic  ordering.
Even (odd) spins fluctuate around $\vphi^{\mathrm{e}}=(0,0,1)\  
(\vphi^{\mathrm{o}}=(0,0,-1)$).
\item{FI:} The ordering consists on even spins fluctuating around the
Z axis and odd spins fluctuating around the cone of angle
$\theta<\pi/2$ with axis Z.
\item{AFI:}
This ordering is similar to the previous one, with $\theta>\pi/2$.
\item{RP$^2$:} Here the ordering is the finite $T$ version of the one
found analytically in $J=-0.5, T=0$, i.e., even spins fluctuating around the Z
axis, with random sense, and odd spins fluctuating around the cone 
with random sense.
\end{itemize}

\subsection{Order Parameters}\label{ORDERPARAMETERS}
In models with antiferromagnetic couplings, one might expect an
even-odd structure of the ordered phases. Therefore, from the local
field $\{\vphi_i\}$, we define the standard magnetization as the
Fourier transform at momentum $0$, and the staggered magnetization as
the Fourier transform at momentum $(\pi,\pi,\pi)$:

\begin{eqnarray}
\M&=&\frac1{L^3}\sum_i\vphi_i\,,\\
\Ms&=&\frac1{L^3}\sum_i(-1)^{x_i+y_i+z_i}\vphi_i\,.
\end{eqnarray}

In a finite lattice we must take the modulus before taking the mean value.
We will study:
\begin{eqnarray}
\muV&=&\langle\Vert\M\Vert\rangle\,,\\
\muVs&=&\langle\Vert\Ms\Vert\rangle\,.
\end{eqnarray}
The associated susceptibilities are
\begin{eqnarray}
\chiV&=&L^3\langle\M^2\rangle\,,\\
\chiVs&=&L^3\langle\Ms^2\rangle\,.
\end{eqnarray}

In order to explore RP$^2$ type phases we introduce the tensor invariant 
under the local spin reversal. In this case we use as local field the 
matrices  $\{\tau_i\}$, constructed as
\begin{equation}
\tau_i=\phi_i^\alpha\phi_i^\beta-\frac13\delta^{\alpha\beta};\ 
\alpha,\beta=1,2,3\,.
\end{equation}
Notice that they are traceless, thus they represent objects of spin 2.
We can now define the associated traceless tensor magnetizations
\begin{eqnarray}
\MT&=&\frac1{L^3}\sum_i \tau_i\,,\\
\MTs&=&\frac1{L^3}\sum_i (-1)^{x_i+y_i+z_i}\tau_i\,,
\end{eqnarray}
and the mean values:
\begin{eqnarray}
\muT&=&\left\langle\sqrt{\mathrm{Tr}\,\MT^2}\right\rangle\,,\\
\muTs&=&\left\langle\sqrt{\mathrm{Tr}\,\MTs^2}\right\rangle\,.
\end{eqnarray}
The corresponding susceptibilities are
\begin{eqnarray}
\chiT&=&L^3\left\langle\mathrm{Tr}\,\MT^2\right\rangle\,,\\
\chiTs&=&L^3\left\langle\mathrm{Tr}\,\MTs^2\right\rangle\,.
\end{eqnarray}

Let us close this subsection recalling the value of the order
parameters (in the large volume limit) in each of the
ordered phases found in the previous subsection:
\begin{equation}
\setlength{\extrarowheight}{4pt}
\begin{array}{lll}
\mathrm{FM}:&\muV>0,\ \muVs=0  &(\Rightarrow\muT> 0,\ \muTs=0)\,,\\
\mathrm{AFM}:&\muVs>0,\ \muV=0  &(\Rightarrow\muT> 0,\ \muTs=0)\,,\\
\mathrm{FI}:& \muV>\muVs>0 &(\Rightarrow\muT,\ \muTs> 0)\,,\\
\mathrm{AFI}:&\muVs>\muV>0 &(\Rightarrow\muT,\ \muTs> 0)\,,\\
\mathrm{RP}^2:&\multicolumn{2}{l}{\muTs>\muT>0,\ \muVs=\muV=0\,.}
\end{array}\label{ORDERPARVAL}
\end{equation}

\subsection{Correlation Length}

For an antiferromagnetic model, not only the usual susceptibility, also
the staggered susceptibility can diverge. Thus, in the Brillouin zone, one needs
to monitor the behavior of the Green functions close to the origin as
well as close to $(\pi,\pi,\pi)$. Since in critical-phenomena studies
one usually considers only the behavior around zero momentum, we have
found it convenient to define four Green functions in terms of four
fields in momentum space
\begin{eqnarray}
\hat{\vphi}(\bm p)&=&\sum_{i} 
\ee^{-{\mathrm i}  \bm p\cdot \bm r_i}\ \vphi_i\ ,\  \\
\hat{\vphi}_{\mathrm s}(\bm p)&=&\sum_{i} 
\ee^{-{\mathrm i}\bm p\cdot \bm r_i}(-1)^{x_i+y_i+z_i}\vphi_i\ ,\\
\hat{\mathsf T}(\bm p)&=&\sum_{i} 
\ee^{-{\mathrm i}  \bm p\cdot \bm r_i}\ \tau_i\ ,\  \\
\hat{\mathsf T}_{\mathrm s}(\bm p)&=&\sum_{i} 
\ee^{-{\mathrm i}  \bm p\cdot \bm r_i}(-1)^{x_i+y_i+z_i}\tau_i
\ ,
\end{eqnarray}
the Fourier transforms of the correlation functions being
\begin{eqnarray}
\hat G^V(\bm p)&=&\frac1{L^3} \langle \hat{\vphi}({\bm p}) \cdot
\hat{\vphi}^*({\bm p})\rangle\,,\label{green1}\\
\hat G^V_\st(\bm p)&=&\frac1{L^3} \langle \hat{\vphi}_\st({\bm p}) \cdot
\hat{\vphi}_\st^*({\bm p})\rangle\,,\label{green2}\\
\hat G^T(\bm p)&=&\frac1{L^3} \langle\mathrm{Tr}\ \hat{\mathsf T}({\bm p})
\hat{\mathsf T}^\dagger(\bm p)\rangle\,,\label{green3}\\
\hat G^T_\st(\bm p)&=&\frac1{L^3} 
\langle\mathrm{Tr}\ \hat{\mathsf T}_\st({\bm p})
\hat{\mathsf T}_\st^\dagger(\bm p)\rangle\,.\label{green4}
\end{eqnarray}

Notice that $\hat G^{V,T}_\st(\bm p)=\hat G^{V,T}(\bm
p+(\pi,\pi,\pi))$, so that one could consider only non staggered
correlation functions that would be studied both close to $(0,0,0)$
and to $(\pi,\pi,\pi)$.

Near a (continuous) phase transition where the corresponding
correlation length, $\xi$, diverges, the correlation functions in the
thermodynamic limit behave for small ${\bm p}^2\xi^2$, as
\begin{equation}
\hat G_(\bm p)\simeq\displaystyle\frac{Z
\xi^{-\eta}}{{\bm p}^2 + \xi^{-2}} \,.
\end{equation}
Here $\xi$ diverges as $|t|^{-\nu}$, $t$ being the reduced
temperature. The anomalous dimension, $\eta$, will depend on the
considered field.

In a finite lattice, to estimate the correlation length one uses the
propagator at zero momentum and at the minimum non-zero momentum
compatible with boundary conditions. Defining $F= \hat G(2\pi/L,0,0)$
and noting that $\chi=\hat G(0)$, one has\cite{COOPER}
\begin{equation}
\xi=\left(\frac{\chi/F-1}{4\sin^2(\pi/L)}\right)^{1/2}\,.
\end{equation}

\section {Mean Field Calculation}\label{MEANFIELDSECT}

When several phases compete, it is quite tricky to calculate the phase
diagram in Mean-Field approximation (the $T=0$ calculation has shown
that we should face this problem!). Since one can find different
ordered phases at low temperatures within different Mean-Field
schemes, it is necessary to decide which phase shall be the most stable
one.  We consider that the cleanest way of performing such a
calculation is to use the variational formulation of the Mean-Field
approximation (see for example Ref.~\onlinecite{parisi}), with a
variational family large enough to take into account all the phases found in
the phase diagram. In this way, all the phases compete on the same
grounds and one has an objective criterion to decide which phase is to
be found in a given region of the phase-diagram.

One needs to compare the actual system with a simplified model where
all degrees of freedom are statistically independent. The method is
derived from the inequality\cite{parisi}
\begin{equation}
F\leq F_0+\langle H-H_0\rangle_0\,.
\label{variational_ineq}
\end{equation}
Here, $H_0$ is a trial Hamiltonian depending on some parameters (the
{\em mean fields}) and the average $\langle \dots \rangle_0$ means
average with the Boltzmann weight corresponding to $H_0$. The
right-hand side of the inequality (\ref{variational_ineq}) is
minimized with respect to the free parameters in $H_0$, then used as
our best estimate of the Free-energy.  Thus the task is to generalize
the standard Curie-Weiss ansatz, $H_0= h \sum_i \phi^z_i$, ($\phi^z_i$
is the component of the local spin $\vphi_i$ along the Z-axis), to
cover all the expected orderings.

In our case, we must use the simplest possible variational family that
permits to have different orderings in the even and odd sublattices:
\begin{equation}
H_0= - \sum_{i\ \mathrm{even}} V_\ee (\phi^z_i) 
- \sum_{i\ \mathrm{odd}} V_\oo (\phi^z_i)\,.
\end{equation}
Notice that, as far as the calculation of the $\langle\ldots\rangle_0$
averages are concerned, all spins can be considered as statistically
independent. Thus, the mean value of an arbitrary function of a spin
placed in (say) the odd sublattice is simply
\begin{equation}
\left\langle f({\vphi} )\right\rangle_0^{(\mathrm{odd})}=
\frac{\displaystyle
\int_0^{2\pi}\,\dd\varphi\  \int_{-1}^{1}\,\dd\phi^z\ f({\vec \phi})
\,\ee^{-V_\oo(\phi^z)/T}}
{\displaystyle\int_0^{2\pi}\,\dd\varphi\  \int_{-1}^{1}\, \dd\phi^z\ 
\,\ee^{-V_\oo(\phi^z)/T}}\,,
\end{equation}
\begin{equation}
\vphi=(\sqrt{1-(\phi^z)^2}\cos\varphi,
\sqrt{1-(\phi^z)^2}\sin\varphi,\phi^z)\,.
\end{equation}

We now need to parametrize the local potentials with the help of the
mean fields, that will be our minimizing parameters.  One easily sees
that keeping only the linear term ($V_{\ee,\oo}(\phi^z)=h_{\ee,\oo}
\phi^z$) will not reproduce the ferrimagnetic or antiferrimagnetic
phases, since at very low temperatures and non vanishing mean-fields,
$h_{\ee,\oo}$, the spins would always be (anti)aligned with the Z
axis. If one keeps also the quadratic term, $V_{\ee,\oo}(\phi^z)=h_{\ee,\oo}
\phi^z + \lambda_{\ee,\oo} (\phi^z)^2$, the situation improves significantly.
The minimum of $V_{\ee,\oo}$ can now be $-1\le \phi^z_\mathrm{min}\le 1$
which implies that at low temperature spins would lie on the cone of
angle $\theta$, $\cos \theta=\phi^z_\mathrm{min}$. Therefore, we will
choose as our variational family
\begin{equation}
\begin{array}{rcl}
H_0&=&\displaystyle-\sum_{i\ \mathrm{even}}
(h_\ee \phi^{z}_i + \lambda_\ee (\phi^{z}_i)^2) \\\\
&&\displaystyle-\sum_{i\ \mathrm{odd}}
(h_\oo \phi^{z}_i + \lambda_\oo (\phi^{z}_i)^2 )\,.
\label{H0}
\end{array}
\end{equation} 
As an extra-bonus, we find that the RP$^2$-phase can be represented by
this ansatz if the Mean Fields that minimize the r.h.s. of inequality
(\ref{variational_ineq}) ---at those particular $T$ and $J$ values--- happen
to be $h_\ee = h_\oo=0$, $\lambda_\ee=-\lambda_\oo >0$. This can be
explicitly checked by calculating the order parameters as a function
of the mean-fields. Due to the symmetry between the even and
odd sublattices, the expressions simplify in terms of the
natural linear combinations of the Mean-Fields
$h_\ee,h_\oo,\lambda_\ee,\lambda_\oo$:
\begin{equation}
\begin{array}{lcl}
h&=&(h_\ee+h_\oo)/2\,,\\
h_\mathrm{s}&=&(h_\ee-h_\oo)/2\,,\\
\lambda&=&(\lambda_\ee+\lambda_\oo)/2\,,\\
\lambda_\mathrm{s}&=&(\lambda_\ee-\lambda_\oo)/2\,.
\label{variables}
\end{array}
\end{equation}
In terms of these variables, by means of a series expansion in $h$,
$h_\mathrm{s}$, $\lambda$ and $\lambda_\mathrm{s}$, 
one gets for the order parameters:
\begin{eqnarray}
\muV&=&\frac{1}{2}(\langle \phi^z\rangle_0^{(\mathrm{even})} + \langle \phi^z
\rangle_0^{(\mathrm{odd})})\\
&=&\frac{2}{3}\beta
h+\frac{8}{45}(h\lambda+h_\mathrm{s}\lambda_\mathrm{s})+
O(h^2,h_\mathrm{s}^2,\lambda^2,\lambda_\mathrm{s}^2)\,,\nonumber
\end{eqnarray}
\begin{eqnarray}
 \muVs&=&\frac{1}{2}(\langle \phi^z \rangle_0^{(\mathrm{even})} 
- \langle \phi^z \rangle_0^{(\mathrm{odd})})\\
&=&\frac{2}{3}\beta h_\st+\frac{8}{45}(h_\mathrm{s}\lambda+h\lambda_\mathrm{s})
+O(h^2,h_\mathrm{s}^2,\lambda^2,\lambda_\mathrm{s}^2)\,,\nonumber
\end{eqnarray}
\begin{eqnarray}
 \muT&=&\frac{1}{2}\left(\langle (\phi^z)^2 \rangle_0^{(\mathrm{even})} 
+ \langle (\phi^z)^2 \rangle_0^{(\mathrm{odd})}\right)-
\frac{1}{3}\\
&=&\frac{4}{45} \beta \lambda +\frac{2}{45}\beta^2 (h^2+h_\st^2)+
\frac{4}{945}\beta^2 (\lambda^2+\lambda_\st^2)\nonumber\\
&&+O(h^2,h_\st^2,\lambda^2,\lambda_\st^2)\,,\nonumber
\end{eqnarray}
\begin{eqnarray}
\muTs
&=&\frac{1}{2}\left(\langle (\phi^z)^2 \rangle_0^{(\mathrm{even})}-\langle
(\phi^z)^2\rangle_0^{(\mathrm{odd})}\right)\\
&=&\frac{4}{45} \beta \lambda_\st\ +\ \frac{1}{45}\beta^2 h h_\st\ +\ 
\frac{8}{945}\beta^2 \lambda \lambda_\st\nonumber\\
&&+O(h^2,h_\st^2,\lambda^2,\lambda_\st^2)\,.\nonumber
\end{eqnarray}
With this information in hand one can identify the different phases
that we found at $T=0$ in terms of the non-vanishing Mean-Fields
(of course the high-temperature PM phase corresponds to the vanishing
of all four mean-fields!): 
\begin{equation}
\begin{array}{lrcllllllll}
\mathrm{FM}:& h&>&0, &h_\st&=&\lambda&=&\lambda_\st&=&0\,,\\
\mathrm{AFM}:& h_\st&>&0, &h&=&\lambda&=&\lambda_\st&=&0\,,\\
\mathrm{FI}:& h,\lambda_\st&>&0, &&&h_\st&=&\lambda&=&0\,,\\
\mathrm{AFI}:& h_\st,\lambda&>&0, &&&h&=&\lambda_\st&=&0\,,\\
\mathrm{RP}^2:& \lambda_\st&>&0, &h&=&h_\st&=&\lambda&=&0\,.
\end{array}
\end{equation}

Let us now describe the actual calculation. As previously said, 
we introduce the function
\begin{equation}
\Phi (h,h_\st,\lambda,\lambda_\st)= F_0+\langle H-H_0\rangle_0\,,
\label{Phi} 
\end{equation} 
that, at its minimum as a function of $h,h_\st,\lambda$ and
$\lambda_\st,$ we shall identify (in Mean-Field approximation) with the
equilibrium free-energy. The partition function can be factorized to
the contribution of the $V/2$ points of the even  sublattice and the
$V/2$ points of the odd sublattice:
\begin{equation}
Z_0=Z_\ee^{V/2} Z_\oo^{V/2}= \ee^{-\beta F_0}\,,
\label{z0}
\end{equation}
where
\begin{eqnarray}
Z_{\ee,\oo}&=&\int_0^{2\pi} \dd \varphi \int_{-1}^1 \dd\phi^z 
\,\ee^{\beta(h_{\ee,\oo}\phi^z +\lambda_{\ee,\oo} (\phi^z)^2)} \,, \\
F_0&=&-\frac{V}{2 \beta} (\log Z_\ee + \log Z_\oo)\,.
\end{eqnarray}

The average of the Mean Field Hamiltonian is
\begin{eqnarray}
\langle H_0 \rangle_0 &=\displaystyle-\frac{V}{2} &\left[h_\ee 
\langle \phi^z\rangle_0^{(\even)} +
\lambda_\ee \langle (\phi^z)^2\rangle_0^{(\even)}\right.\\\nonumber
&&+\left.h_\oo \langle \phi^z\rangle_0^{(\odd)} +
\lambda_\oo \langle (\phi^z)^2\rangle_0^{(\odd)}\right]\,.
\label{H0med}
\end{eqnarray}
As for the average of the true Hamiltonian, one finds:
\begin{eqnarray}
\langle H\rangle_0&=&- 3VJ \langle 
\phi^z\rangle_0^{(\even)}\langle \phi^z\rangle_0^{(\odd)}\\\nonumber
&&- 3VJ \left\langle \sqrt{1+\vphi_\ee\cdot\vphi_\oo} \right\rangle_0\,.
\label{HPROBLEM}
\end{eqnarray}
In the above expression, $\vphi_{\ee,\oo}$ is a generic spin
belonging to the even (odd) sublattice.
The problem is that, even if $\vphi_\ee$ and
$\vphi_\oo$ are statistically independent, the calculation of the
mean-value of the square root in Eq.~(\ref{HPROBLEM}) cannot be
straightforwardly factorized in an even and an odd
contribution. In order to achieve this factorization, we shall use the
series expansions introduced by de Gennes.\cite{DeGennes} One first
use an expansion in Legendre Polynomials:
\begin{equation}
\sqrt{1+\vphi_\ee\cdot\vphi_\oo}= \sum_{l=0}^{\infty}A_l
P_l\left(\vphi_\ee\cdot\vphi_\oo\right)\,,
\label{SERIEPOL}
\end{equation}
\begin{equation}
A_l=(-1)^{l+1}\frac{2\sqrt{2}}{(2l-1)(2l+3)}\,.
\end{equation}
We can now factorize the Legendre Polynomials using their expression in
terms of spherical harmonics
\begin{equation}
P_l(\vphi_\ee\cdot\vphi_\oo)=\frac{4\pi}{2l+1}\sum_{m=-l}^l 
Y_l^{m*}(\phi^z_\ee,\varphi_\ee)Y_l^{m}(\phi^z_\oo,\varphi_\oo)\,.
\label{HARMONIC}
\end{equation}
Thus, the mean-values are factorized into even and odd contributions.
Due to the rotational symmetry along the Z axis, only the $m=0$ terms
in Eq.~(\ref{HARMONIC}) are non 
vanishing. Thus we obtain,
\begin{equation}
\langle P_l(\vphi_\ee\cdot\vphi_\oo)\rangle_0
=\langle P_l(\phi^z)\rangle_0^{(\even)}
 \langle P_l(\phi^z)\rangle_0^{(\odd)}\,.
\label{prodP}
\end{equation}
Fortunately, if one wants to calculate the free-energy $\Phi$ as a
series expansion in the Mean-Fields, $h$, $h_\st$, $\lambda$ and
$\lambda_\st$, at a given order only a finite number of terms in
Eq.~(\ref{SERIEPOL}) contribute, due to the orthogonality properties
of the Legendre Polynomials. This expansion allows to discuss the
continuous phase transitions from the PM phase (where
$h=h_\st=\lambda=\lambda_\st=0$ is the absolute minimum of the
free-energy), to ordered phases. Indeed, calculating $\Phi$ (per unit
volume) to second order one gets:
\begin{eqnarray}
\frac1V\,\Phi(h,h_\st,\lambda,\lambda_\st)&\approx 
&(\frac{\beta}{6}-\frac{J\beta^2}{3}-\frac{2\sqrt{2}\beta^2}{15})h^2\\\nonumber
&&+(\frac{\beta}{6}+\frac{J\beta^2}{3}+
\frac{2\sqrt{2}\beta^2 }{15})h_\st^2\\\nonumber
&&+(\frac{8\sqrt{2}\beta^2}{1575}+\frac{2\beta}{45})\lambda^2\\\nonumber
&&+(-\frac{8\sqrt{2}\beta^2}{1575}+\frac{2\beta}{45})\lambda_\st^2\,.
\label{QUADRATIC}
\end{eqnarray}
This is a quadratic form in $h$, $h_\st$, $\lambda$ and $\lambda_\st$.  If
the above quadratic form is positive definite, the PM phase is a
(local) minimum of the free energy. The other way around, when one of
the eigenvalues of the quadratic form is negative, the PM phase is
unstable with respect to some ordered phase, depending on the
Mean-Field that should grow in order to minimize the
free-energy. Notice also that the eigenvalue corresponding to
$\lambda^2$ is always positive.  Thus, even if there are four
eigenvalues, we obtain three lines of continuous phase transitions,
where the eigenvalues vanish:
\begin{equation}
\begin{array}{ll}
\hbox{PM-FM line:} &T=2J + 4\sqrt{2}/5\,,\\
\hbox{PM-AFM line:} &T=-2J - 4\sqrt{2}/5\,,\\
\hbox{PM-RP2 line:} &T=4\sqrt{2}/35\,.
\end{array}
\end{equation}

\begin{figure}
\includegraphics[angle=90,width=\columnwidth]{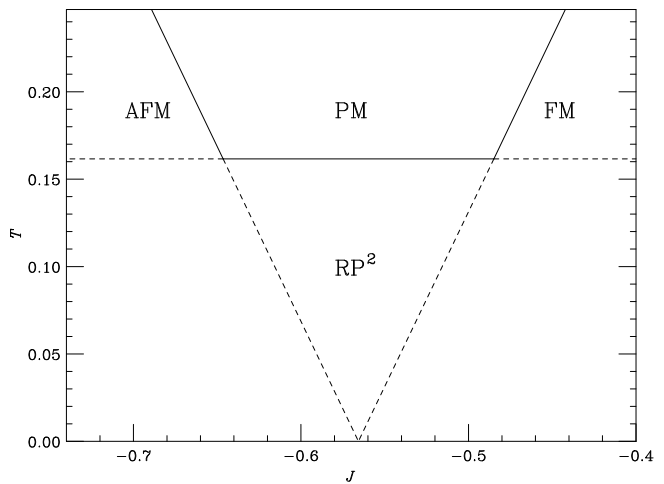}
\caption{Phase diagram as obtained from the second order series
expansion of the free-energy (\protect{\ref{QUADRATIC}}). The
paramagnetic phase is unstable for temperatures below the full-lines
(the instability being toward the FM, AFM or RP$^2$ phases, as
indicated in the plot). The dashed lines indicate the places where
some of the eigenvalues of the quadratic form in
(\protect{\ref{QUADRATIC}}) vanish, but they do not correspond to
phase transitions.  }
\label{ORDEN2}
\end{figure}

Therefore the PM phase, stable at high-temperature, meets two transition
lines of opposite slope, and an horizontal line that separates it
from the RP2 phase (see Fig.~\ref{ORDEN2}). 

For temperatures below the full lines in Fig.~\ref{ORDEN2}, one needs
to discuss the stability of a minimum of the free-energy different
from $h=h_\st=\lambda=\lambda_\st=0$. To locate that minimum, and to
discuss its stability, one needs to extend the series expansion in
(\ref{QUADRATIC}) at least to fourth-order in $h$, $h_\st$, $\lambda$
and $\lambda_\st$. This can be done (see Appendix~\ref{FOURTH}), but
it is not particularly illuminating since the series expansion for
$\Phi$ is not fastly convergent. We have rather turned to a numerical
method.  Given a particular value of the Mean-Fields, $h$, $h_\st$,
$\lambda$, $\lambda_\st$, we have calculated $\Phi$ by means of a
Gauss-Legendre integration of all the terms in Eq.~(\ref{Phi}). To do
this, we have divided the interval $[-1,1]$ in twelve subintervals and
we have done a twelfth order Gauss-Legendre integration in each of
them. The series of Eq.~(\ref{SERIEPOL}) has been evaluated to order
50. Being able to calculate $\Phi$, the minimization has been done
using a conjugate gradient method. The resulting phase diagram is
shown together with the results of the Monte Carlo calculation in
Fig.~\ref{diag}.  Notice that, as usual, the Mean-Field
calculation overestimates the critical temperatures by a large factor
(about 2.3 in this case). However, considering this rescaling of the
temperatures, the obtained phase-diagram is in remarkable agreement
with the numerical one.  We have confirmed that all the transitions
are second order except the ferromagnetic-RP2 which is first order
(nevertheless, this transition line is an artifact of the Mean-Field
solution: in the Monte Carlo phase-diagram it seems to collapse to a
tetracritical point). The second order nature of the transitions can
be checked by computing the appropriate order parameter at a given
value of $T$ and $J$, then noticing that it vanishes at the transition
line with Mean-Field exponents ($M\propto |T-T_{\mathrm{c}}|^{1/2}$ or
$\propto |J-J_{\mathrm{c}}|^{1/2}$).

Since Mean-Field overestimates critical temperatures, it is
interesting to compare the previous results with the ones of another
approximation (large $N$) that usually underestimates them. We have
calculated the position of the PM-FM and PM-AFM phase-transition in
the large $N$ approximations (see Appendix~\ref{LARGEN}):
\begin{equation}
\begin{array}{ll}
\hbox{PM-FM line:} &T=+1.2578\,J+0.5578\,,\\
\hbox{PM-AFM line:} &T=-1.2578\,J-0.793\,.
\end{array}
\end{equation}
The critical temperature is underestimated by roughly the same factor
that the Mean Field approximation overestimates it (see below). To
extend further this calculations would require to study non
traslationally invariant saddle-points, which is rather complex.

\section{Monte Carlo simulation}\label{MONTECARLOSECT}

The model (\ref{accion}) can be investigated using a standard Monte
Carlo method. We shall here describe some technical points, the
results being discussed in the following subsections.

The single Monte Carlo (MC) step consists of a full-lattice Metropolis
lattice sweep. Some of the simulations have been done at extremely low
temperatures, thus the method of choice would have been a heat-bath
algorithm, but its implementation in this model is rather complex. 
Fortunately, one can effectively falsify a heat-bath
algorithm by means of a multi-hit Metropolis method, proposing per each hit as
spin update a random spin on the unit sphere.  Luckily enough, to
achieve a 50\% acceptance the number of needed hits is quite modest
except for the lowest temperatures that represent a negligible
fraction of the total CPU time devoted to the problem.  The
pseudo-random number generator was the Congruential+Parisi-Rapuano
(see e.g.  Ref.~\onlinecite{SDHV}).

To extract critical exponents and critical temperatures, we have used
the quotient methods:\cite{RP2D3,cocientes2} for a pair of lattices of
sizes $L$ and 2$L$ we choose the temperature where the
correlation-lengths in units of the lattice-size coincide
($2\xi_L =\xi_{2L}$). Up to scaling corrections, the matching
temperature is the critical point. Let now $O$ be a generic observable
diverging at the critical point like $|t|^{-x_O}$. Then, one has (up
to scaling corrections\cite{RP2D3,cocientes2}):
\begin{equation}
\left.\frac{\langle O\rangle_{2L}}{\langle
O\rangle_{L}}\right|_{\frac{\xi_{2L}}{\xi_L}=2}=2^{x_O/\nu}\,,
\label{QUO}
\end{equation}
where $\nu$ is the critical exponent for the correlation length
itself.  For extracting $\nu$ we have used the temperature derivative
of the correlation-length, $x_{\partial_T \xi}=1+\nu$. To fulfill the
matching condition $2\xi_L =\xi_{2L}$ one often needs to extrapolate
from the simulation temperature to a nearby one. This has been done
using a reweighting method (see e.g. Ref.~\onlinecite{FERRSWEN}).

\subsection{Phase Diagram \label{PHDI}}

\begin{figure}
\includegraphics[angle=90,width=\columnwidth]{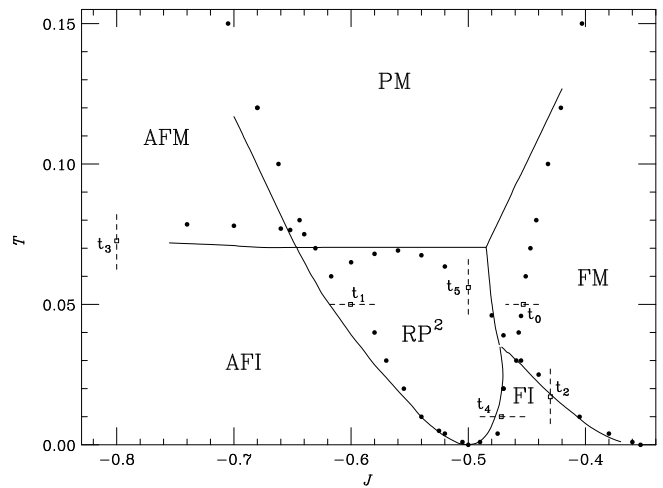}
\caption{Phase diagram of the model. Circles correspond to Monte Carlo
estimates. Full lines correspond to the Mean-Field transition
temperatures divided by 2.3. The dashed vertical and horizontal
lines show the direction of the Ferrenberg-Swendsen extrapolation in
the points which are the subject of detailed study reported. We label
t$_0\dots$t$_5$ each one of the more carefully  studied transitions.}
\label{diag}
\end{figure}

At a first stage, the phase diagram (see Fig.~\ref{diag}) has been
explored by performing hysteresis cycles with $L$ from $6$ to
$24$. The energy and relevant order parameters have been used to
locate phase transitions. Once a rough diagram has been established,
we have simulated in lattices $L=6,12$, and have refined the
transition point calculating the correlation lengths associated to the
relevant order parameter and computing the point where $\xi/L$ is $L$
independent.  The phase diagram obtained can be seen in
Fig.~\ref{diag}.  We have studied in greater detail the phase
transitions at the points t$_0$,\ldots, t$_5$, depicted in
Fig.~\ref{diag}. We have simulated lattices $L=6,8,12,16,24,32,48$ and
$64$, producing 20 million of MC full-lattice sweeps for the largest
lattices in each transition.  We have discarded $5\times 10^5$ MC
steps for thermalization. In all cases this has been checked to be
much larger that the integrated autocorrelation time. In addition, at
the lowest temperatures, we have compared different starting
configurations (random, FM, etc.), concluding that the results are
start-independent.

\begin{table*}
\begin{ruledtabular}
\begin{tabular}{lcccccc}
Transition& $L=6$&$L=8$&$L=12$&$L=16$&$L=24$&$L=32$  \\
\hline
t$_0$ ($T=0.05,\muV$) 
&-0.453561(15)&-0.453293(32)&-0.453131(19)&-0.453090(15)&-0.453091(29)&--- \\
\hline
t$_1$ ($T=0.05,\muV$) 
&-0.59828(8) &-0.59939(4) &-0.60015(2) &-0.60038(1) &-0.60043(2)&-0.60044(2) \\
\hline
t$_1$ ($T=0.05,\muVs$) 
&-0.60083(4) &-0.60084(3) &-0.60078(2) &-0.60067(1) &-0.60052(2)&-0.60048(2) \\
\hline
t$_2$ ($J=-0.43,\muVs$) 
&0.017663(12)& 0.017343(5)&0.017163(4)&0.017129(2)&0.017112(2)&0.017101(4)\\
\hline 
t$_3$ ($J=-0.8,\muV$) 
&0.07528(4) &0.07387(2) &0.07304(2) &0.07283(1) &0.07267(1)&0.07260(1) \\
\hline
t$_4$ ($T=0.01,\muV$) 
&-0.47199(3) &-0.47198(2) &-0.47196(2)&-0.47195(1)&-0.471919(6)&-0.471916(3) \\
\hline
t$_4$ ($T=0.01,\muVs$) 
&-0.47241(3) &-0.47219(3) &-0.47201(2)&-0.47196(1)&-0.471914(6)&-0.471912(3) \\
\end{tabular}
\end{ruledtabular}
\caption{$J_\mathrm{c}$ or $T_\mathrm{c}$ determined by the intersection of the correlation 
lengths measured in two lattices of size $L$ and $2L$. t$_N(X,A)$ stands for:
t$_N$, transition, $X$ fixed parameter and $A$ the order parameter associated
with the correlation length considered.} 
\label{punto_critico}
\end{table*}

\begin{table*}
\begin{ruledtabular}
\begin{tabular}{lcccccc}
Transition & $L=6$&$L=8$&$L=12$&$L=16$&$L=24$&$L=32$  \\
\hline
t$_0$ ($T=0.05,\muV$) 
&0.707(4) &0.702(7) &0.712(12) &0.710(10) &0.629(95)&--- \\
\hline
t$_1$ ($T=0.05,\muV$) 
&0.594(20) &0.556(7) &0.555(8) &0.540(8) &0.546(17)&0.596(25) \\
\hline
t$_1$ ($T=0.05,\muVs$) 
&0.592(5) &0.561(6) &0.538(5) &0.519(7) &0.517(13)&0.561(22) \\
\hline
t$_2$ ($J=-0.43,\muVs$) 
&0.591(8) &0.569(5) &0.537(3) &0.548(5) &0.588(8)&0.604(17) \\
\hline 
t$_3$ ($J=-0.8,\muV$) 
&0.583(10) &0.557(4) &0.562(3) &0.582(6) &0.605(7)&0.651(20) \\
\hline
t$_4$ ($T=0.01,\muV$) 
&0.534(4) &0.536(10) &0.560(10) &0.597(15) &0.630(17)&0.656(24) \\
\hline
t$_4$ ($T=0.01,\muVs$) 
&0.545(7) &0.564(13) &0.581(13) &0.611(16) &0.629(17)&0.650(25)
\end{tabular}
\caption{Apparent $\nu$ exponent obtained from the quotient method applied to 
$(L,2L)$ pairs.}
\label{nus}
\end{ruledtabular}
\end{table*}

Before discussing the results let us briefly comment what can one
expect on universality grounds. Transition t$_0$ connects the
paramagnetic phase, where the full O(3) symmetry-group is preserved, to
a FM phase where the symmetry group is just the O(2) group
corresponding to the global rotations around the global magnetization.
Thus it is expected (and confirmed) to be in the Universality class of
the O(3) Non Linear $\sigma$ model (see table \ref{exp_ref}). For all
the other transitions the scheme of symmetry breaking is not so clear.
The only obvious symmetry-breaking (for transitions t$_2$ and t$_3$)
is the symmetry between the even and odd sublattice. This is a Z$_2$
symmetry, thus one might expect the transition to be in the
Ising Universality class. The symmetries of the RP$^2$ phase are
intriguing and will be investigated in the following
subsection. Let us only recall that the transition t$_5$ has been
recently studied in great detail in Ref.~\onlinecite{RAIZLETTER}, 
where it was found that
\begin{eqnarray}
T_\mathrm{c}&=&0.055895(5)\,,\label{tcrp2}\\
\nu&=&0.781(18)\,.
\end{eqnarray}
Perhaps not unexpectedly, the critical exponents were found to be
compatible within errors with the one of the antiferromagnetic RP$^2$
model.~\cite{RP2D3}

\begin{table}[h]
\begin{ruledtabular}
\begin{tabular}{cc}
Model        & $\nu$        \\\hline
O(1) (Ising)~\cite{ISPERC} & $0.6294(10)$ \\\hline
O(2)~\cite{XYCLASS}         & $0.67155(27)$\\\hline
O(3)~\cite{HASENBUSCH}         & $0.710(2)$   \\\hline
O(4)~\cite{HASENBUSCH}         & $0.749(2)$   \\\hline
RP$^2$-AFM~\cite{RP2D3}    & $0.783(11)$ \\\hline
Chiral(Heisenberg)~\cite{ChiralN23}  & $0.57(3)$    \\\hline
Chiral(XY) ~\cite{ChiralN23} & $0.55(3)$ \\\hline
Tricritical~\cite{AMIT}  & $1/2     $ \\\hline
Weak 1$^\mathrm{st}$ order~\cite{LOMBARDO} & $1/2$  \\\hline
1$^\mathrm{st}$ order& $1/3$
\end{tabular}
\caption{Critical exponent $\nu$ for some three dimensional Universality
Classes.}
\label{exp_ref}
\end{ruledtabular}
\end{table}

\begin{figure}[ht]
\includegraphics[angle=90,width=\columnwidth]{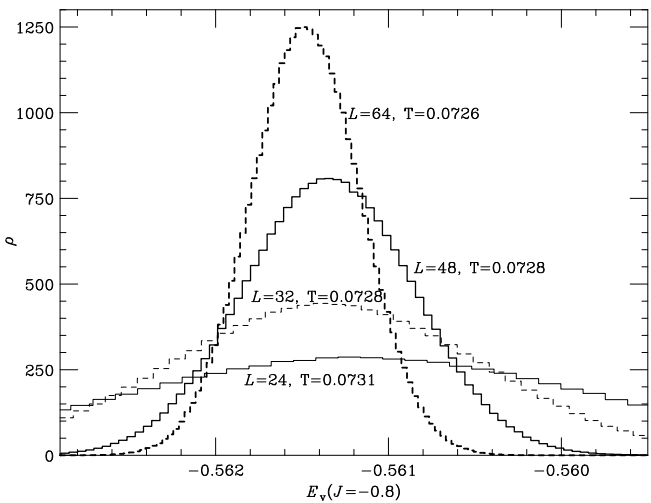}
\caption{Histogram for $E_\mathrm{v}=\langle\vphi_i\cdot\vphi_j\rangle$ 
(for nearest neighbors $i$ and $j$), in transition t$_3$.}
\label{frec1}
\end{figure}

As for transitions t$_0$, \ldots, t$_4$, we have located quite
accurately the critical parameters (see table~\ref{punto_critico}). We
have focused in each case in the largest order parameter (see Eq.~(\ref{ORDERPARVAL})). The PM-AFM transition should have
the same critical behavior and we have not invested time in this
study.  We are reasonably confident in the continuous nature of all
five transitions. This stems from two facts. First, the energy
histograms are not double-peaked (see an example in
Fig.~\ref{frec1}). Yet, a much more refined test comes from the
($L$-dependent) value of the effective $\nu$ exponents shown in
table~\ref{nus}. With the exception of transition t$_0$, which as
expected belongs to the Universality Class of the Heisenberg model in
three dimensions, scaling-corrections are not even monotonous in their
evolution with the lattice size. Although an asymptotic value can not
be guessed with reachable lattice sizes, at least one sees that, for
the largest lattices, the exponent $\nu$ is reasonably far from the
value $1/2$ to be expected in weak first order transitions.

\subsection{Detailed study of the RP$^2$ phase} \label{RP2SUBSECT}

The RP$^2$ phase poses many questions,the first one possibly being: is
it truly a RP$^2$ phase? In other words, (as far as long-distance
correlations are concerned) do the spins behave like {\em segments}
rather than {\em arrows}? As we will see below, the answer is
positive: the spin ordering is truly invariant under spin
reversal. Once this is established, one may worry about continuous
symmetries. At the special point $T=0$, $J=-0.5$, the spins in the
(say) even sublattice are randomly aligned or anti-aligned with the
(say) Z-axis. The spins in the odd sublattice are randomly placed on
the XY-plane. This state has a remaining O(2) symmetry, corresponding
to rotations around the Z-axis. The Z$_2$ symmetry corresponding to
exchanging the roles of the even and odd sublattices is obviously
broken. Hence one may wonder if, up to the critical temperature which
separates it from the paramagnetic state, the RP$^2$ phase can be
characterized as an O(2)-symmetric phase with broken even-odd
symmetry. Doubts arise from the order-from-disorder interaction that
favors the alignment\cite{RP2D3} (in the segment sense) of all the
spins in the planar sublattice (the odd sublattice in the above
discussion). This global alignment would imply a second phase
transition separating the low temperature O(2)-symmetric phase from a
phase without any remaining rotational symmetry at higher
temperature. This second RP$^2$ phase would be separated from the
paramagnetic state where the full rotational symmetry, O(3), is
maintained. This complex scenario would explain the exotic values of
the critical exponents corresponding to the RP$^2$-PM
transitions. Universality arguments\cite{AZARIA} would then predict
that the critical exponents would be the ones of the O(4) classical
Non-Linear $\sigma$ model, which are not very far from the ones
actually found\cite{RAIZLETTER}. On the other hand, in the simple
scenario of a single O(2) symmetric RP$^2$ phase, the expected
critical exponents would be those of the classical Heisenberg model,
in plain contradiction with our numerical results.  Since the simple
scenario seems to be the real one, we are facing a failure of the
Universality Hypothesis for which we do not have any plausible
explanation.

To summarize, in the following we shall address, in this order,
the three following questions:
\begin{enumerate}
\item Is the  RP$^2$ phase truly segment-like?
\item Is the even-odd symmetry broken up to the temperature separating the
RP$^2$ phase from the Paramagnetic state?
\item Is the low temperature O(2) symmetric RP$^2$ phase preserved up
to the temperature separating the RP$^2$ phase from the Paramagnetic
state?
\end{enumerate}

\subsubsection{Tensor versus vector ordering}

\begin{figure}[h!]
\includegraphics[angle=90,width=\columnwidth]{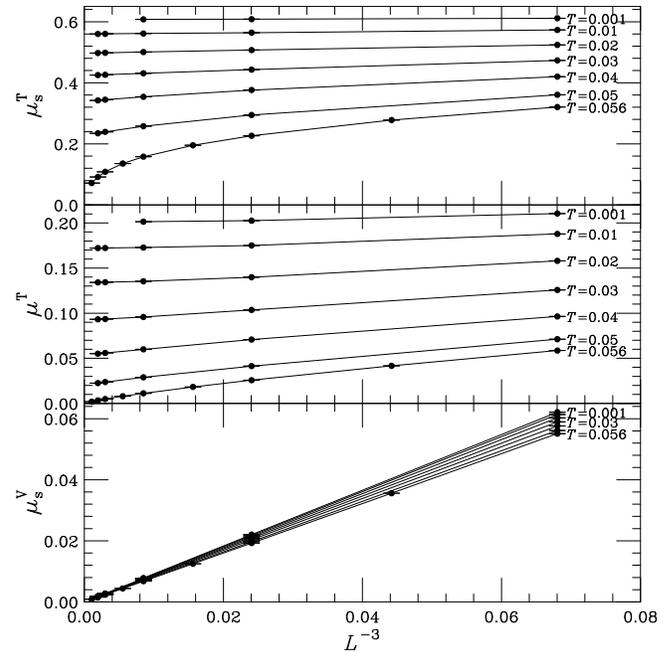}
\caption{Lattice size dependence of $\muTs$, $\muT$ and $\muVs$ at different 
temperatures, for $J=-0.5$.}
\label{LimTer}
\end{figure}

We have called RP$^2$ the phase in which the vector magnetization
vanishes (for any momentum in the Brillouin zone), and the tensor
magnetization is non-vanishing, both at momentum $(0,0,0)$ ($\muT$)
and at momentum $(\pi,\pi,\pi)$ ($\muTs$). In Fig.~\ref{LimTer} (upper
and middle part) we show, fixing $J=-0.5$, that, for temperatures
ranging from 0.001 to 0.05, there is a non vanishing thermodynamic
limit for both quantities. For comparison, we show in the lower part
the vector magnetization at momentum $(\pi,\pi,\pi)$ ($\muVs$), which
goes to zero as $1/\sqrt{L^3}$. We have also measured the
correlation-length in the vector channel. Although some short-range
ordering is present, the correlation-length is not larger than $0.3$
lattice spacing.

\begin{figure}
\includegraphics[angle=90,width=\columnwidth]{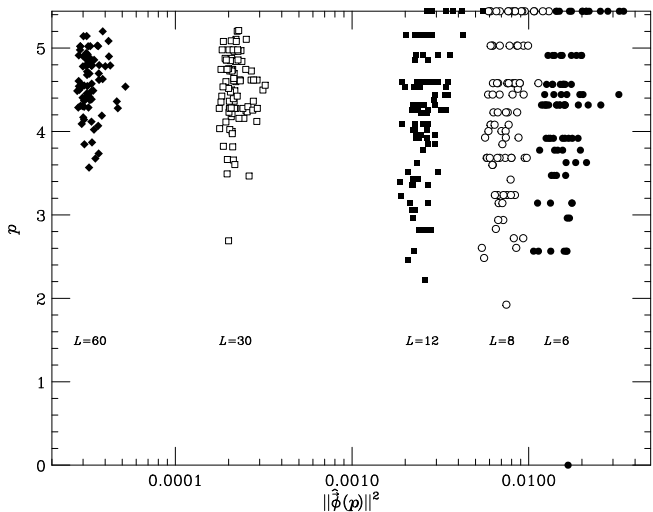}
\caption{Scatter-plot for 90 statistically independent configurations for each  
lattice size ($L=6,8,12,30,60$) at $J=-0.55$, $T=0.5$. In the horizontal axis 
we plot the maximum over the Brillouin zone of the squared Fourier 
transform of the spin field, and in the vertical axis the corresponding 
associated momentum $p=\Vert\bm p\Vert$. The horizontal position of legends 
scales as $L^{-3}\log L$ in agreement with the absence-of-order prediction.}
\label{Fourier}
\end{figure}

To confirm the absence of any other vectorial magnetization we have
measured at $J=-0.55, T=0.5$ (just in the middle of the RP$^2$ phase)
all Fourier components of the vector field $\vphi_i$ for 90
statistically independent configurations for each lattice size, and
plotted in Fig.~\ref{Fourier} the corresponding momentum versus the
maximum value of the Fourier component squared. In other words, we are
searching for the maximum (over the Brillouin zone) of the static
structure factor (divided by $L^3$). We have chosen as lattice sizes
$L=6,8,12,30,60$ to allow for different periodicities of the would-be
vector-ordered states.  If no vectorial ordering is present, the last
quantity should go to zero as $1/L^3$, up to logarithmic corrections
that arise from the fact that we are computing the maximum of a set
of $O(L^3)$ elements.  The absence of ordering is clear from
Fig.~\ref{Fourier}.

\subsubsection{Even-Odd symmetry}

To analyze the even-odd symmetry, we measure the tensor correlation 
difference at 
second neighbors between even and odd lattices. The normalized total difference
for a given configuration can be written as
\begin{equation}
\Delta_{E}=\frac2{3L^3}\left(\sum_\mathrm{even}(\phi_i\cdot\phi_j)^2-\sum_\mathrm{odd}(\phi_i\cdot\phi_j)^2\right)\,,
\end{equation}
where the sums extend over even (odd) second neighbor pairs.  The
non-vanishing of the difference in the thermodynamic limit signals
even-odd symmetry breaking.  Notice that the sublattice energy
difference can be defined locally, and it plays the role of a local
field.  Another interesting observable is the dimensionless quantity
associated to the energy difference
\begin{equation}
\kappa_E=\frac{\langle \Delta_E^2 \rangle}
{\langle \Delta_E \rangle^2}\ .
\end{equation}

\begin{figure}
\includegraphics[angle=90,width=\columnwidth]{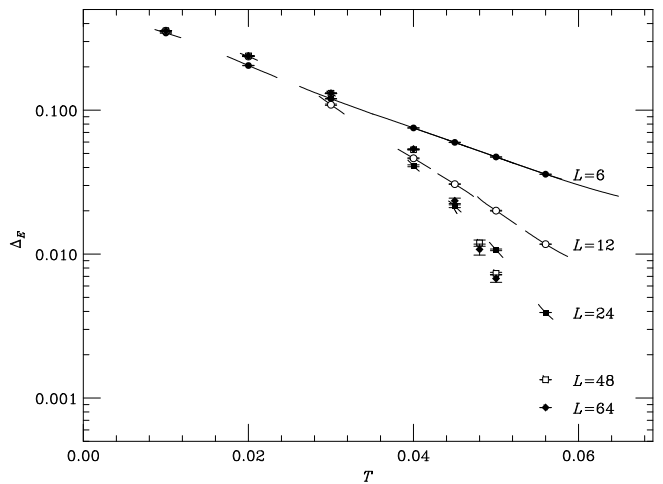}
\caption{Difference of the tensor second neighbors energies between 
sublattices, for $T=0.05$\,.}
\label{MdET}
\end{figure}

\begin{figure}
\includegraphics[angle=90,width=\columnwidth]{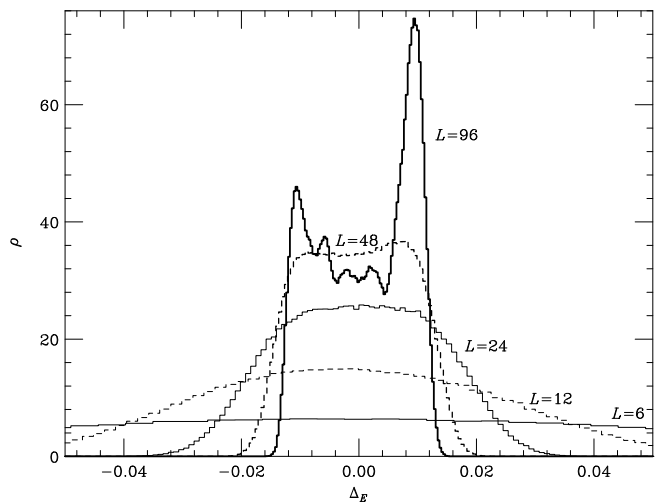}
\caption{Histogram of $\Delta_E$ for $J=0.5$, $T=0.05$\,.}

\label{dEThistog}
\end{figure}

Fig.~\ref{MdET} shows the tensor energy difference as a function of temperature
for several lattice sizes. A clear thermodynamic limit can be observed for
$T<0.05$\ . At $T=0.05$ the asymptotic behavior can be elucidated by a direct
study of the tensor energy difference histograms. A $L=96$ lattice is necessary
to clearly resolve the two peak structure of the histogram 
(see Fig.~\ref{dEThistog}), corresponding to an even-odd symmetry breaking.
Notice that $L=96$ is the largest lattice used, which makes it impossible to 
study the thermodynamic limit of that quantity for $T$ larger than $0.05$ and 
less than $T_\mathrm{c}=0.0559$. We can conclude that, within the computational
resources employed, no evidence exists for a thermodynamic limit with unbroken
even-odd symmetry.

\begin{figure}
\includegraphics[angle=90,width=\columnwidth]{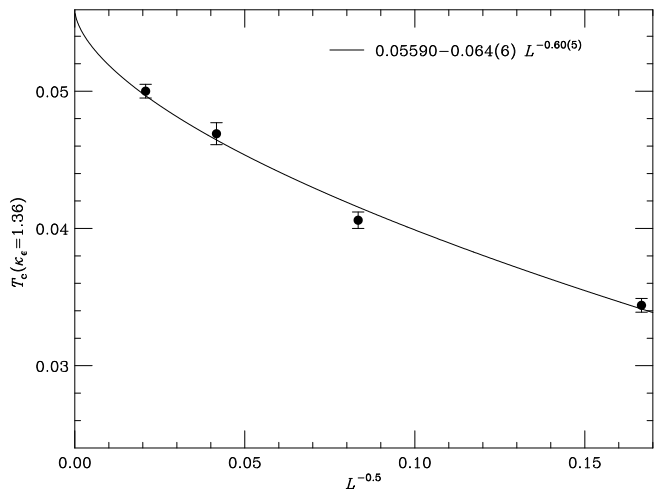}
\caption{Displacement of the critical temperature, defined as the point
where $\kappa_E$
takes a fixed value, as a function of size.
Fits suggest that there is no even-odd symmetry restoration at a 
temperature less than the RP$^2$-PM one.}
\label{BinderfijoE}
\end{figure}

Although no thermodynamic limit can be reached beyond $T=0.05$, more
information can be obtained through a finite size analysis. The closer
we get to $T=0.05$, the harder it becomes to find a two peak structure
in the histogram. A correlation length could be defined in the
even-odd symmetry breaking channel which grows as the possible
critical point between the RP$^2$ phase with broken even-odd symmetry
and a hypothetical RP$^2$ phase with restored even-odd symmetry is
approached. The functional form of the growth of the correlation
length might give indication of the existence of such phase
transition. A direct way to carry out that study is to define the
correlation length as the lattice-size itself, when the histogram has
a central valley at half the peak hight. The result shows a growth of
the correlation length as $T$ increases compatible with a divergence
just at $T_\mathrm{c}$, though with rather peculiar exponents. But the
measurement of that correlation length is very noisy. A much more
precise way to study the possible presence of a transition previous to
$T_\mathrm{c}$ is to define as apparent critical point the $T$ value
at which $\kappa_E$ takes a fixed value. Fig.~\ref{BinderfijoE} shows
the results. Although the possibility of an even-odd symmetry recovery
transition previous to $T_\mathrm{c}$ cannot be discarded, results are
compatible with a divergence just at $T_\mathrm{c}$. The figures show
a fit to a power law (fixing the critical point to the value given in
Eq.~(\ref{tcrp2})). It is worth remarking that the $\nu$ exponent
obtained is very large (2 or larger), which might point to a
logarithmic divergence. Thus, all our results point to a single RP$^2$
phase with broken even-odd symmetries at all temperatures.

\subsubsection{O(2) Symmetry }

The chosen tool to study whether the O(2) symmetry of the $T=0$ state
is preserved at higher temperatures, has been the eigenvalue structure
of the tensor $\MTs$. The latter being traceless implies that the
vector $\boldsymbol{\lambda}=(\lambda_1,\lambda_2,\lambda_3)$ must lie
on the $x+y+z=0$ plane. The whole information reduces, then, to a
modulus (which is nothing but the observable $\mu_{\mathrm s}$), and an
angle, which contains all the information of the eigenvalues on the
symmetry O(2). As any result must be symmetric under eigenvalue
permutations and global inversion, we can restrict the angle to the
interval between 0 and $\pi/6$. More precisely, we consider the
orthonormal basis $\{{\bm u}_x,{\bm u}_y\}$ for the plane given by
\begin{eqnarray}
{\bm u}_x&=&\frac1{\sqrt2}(-1,1,0)\ ,\\
{\bm u}_y&=&\frac1{\sqrt6}(-1,-1,2)\ .
\end{eqnarray}
and define the angle $\theta$ from the relation

\begin{equation}
\tan{\theta}=\frac{{\bm \lambda}\cdot{\bm u}_y}
{{\bm \lambda}\cdot{\bm u}_x}\ ,
\end{equation}
with the proviso that we choose a permutation and a global sign such that
$\theta$ lies between 0 and $\pi/6$.

Another interesting quantity can be defined as follows. In the
thermodynamic limit an O(2)-symmetric phase corresponds to
$\lambda_2=\lambda_3$. We thus define
\begin{equation}
\Delta_\lambda=|\lambda_2-\lambda_3|\ .
\end{equation}
which must vanish in the thermodynamic limit if the O(2) symmetry is 
not broken. As usual the corresponding dimensionless quantity is
\begin{equation}
\kappa_\lambda=\frac{\langle \Delta_\lambda^2 \rangle}
{\langle \Delta_\lambda \rangle^2}\ .
\end{equation}

\begin{figure}
\includegraphics[angle=90,width=\columnwidth]{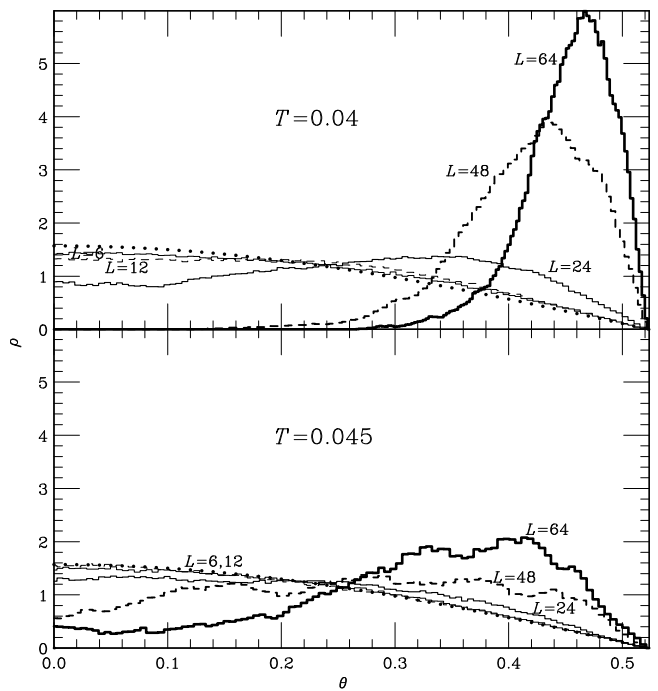}
\caption{Histograms of the angle of the eigenvalue vector on the (1,1,1) plane 
for two temperatures at $J=-0.5$. 
The dots correspond to paramagnetic configurations.}
\label{angulos}
\end{figure}
Fig.~\ref{angulos} shows histograms of angles at several temperatures
and lattice sizes. Dotted lines correspond to completely disordered 
configurations. In case of the system
being O(2)-symmetric (one large eigenvalue and two identical small
eigenvalues), the distribution should be a delta at angle
$\pi/6$. For a system with broken O(2)-symmetry but unbroken
even-odd-symmetry, the eigenvalues are $(a,0,-a)$-like, which would
correspond to a Dirac delta at angle 0. We notice that, for small lattices,
the distribution points to complete disorder, but as the size grows an
inflection point turns up at $T=0.04,0.045$ for $L=24,48$
respectively, and as $L$ goes on growing a peak arises at angles ever
closer to the maximum. It might be said, that the behavior in $L$ is
always the same, except for a scale change.

\begin{figure}
\includegraphics[angle=90,width=\columnwidth]{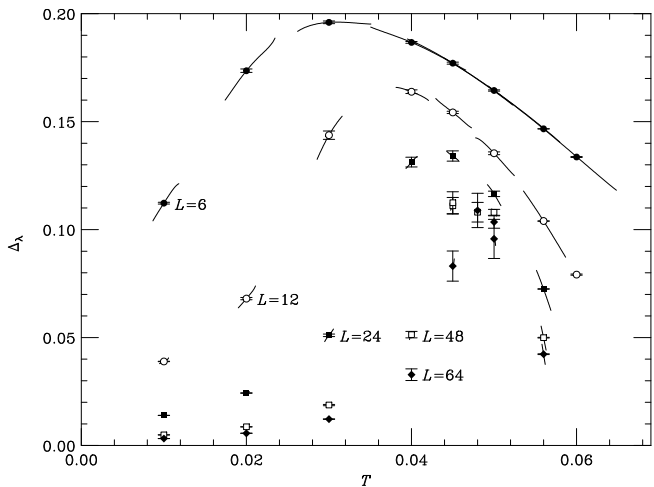}
\caption{Modulus of the difference between the two smaller eigenvalues of 
$\MTs$ as a function of $T$.}
\label{Mdeig}
\end{figure}

Another interesting quantity is the difference between the two small
eigenvalues ($\Delta_\lambda$), which should vanish in the presence of 
O(2) symmetry, 
so turning out to be an order parameter. Fig.~\ref{Mdeig} shows that 
quantity for several values of the temperature and lattice size. If we look
at an intermediate size ($L=24$ for instance) the appearance is that of a 
transition
at $T=0.03$ to a phase with broken O(2) symmetry. Yet, as $L$ increases, 
the apparent transition moves back approaching $T_\mathrm{c}$ ever more.

\begin{figure}
\includegraphics[angle=90,width=\columnwidth]{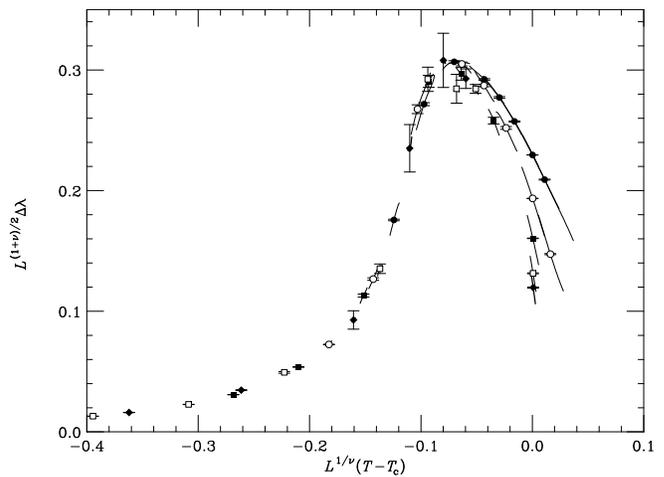}
\caption{Scaling of $\Delta_\lambda$ for the analysis of a 
possible O(2) restoring
transition. Data is fairly well fitted assuming the transition occurs at
$T_\mathrm{c}$ (Points next to $T_\mathrm{c}$ are not well fitted because
there the largest eigenvalue becomes zero and $\Delta_\lambda$ 
ceases to make sense).
The fitted values are $\eta=-0.5$, $\nu=1.8$, $T_\mathrm{c}=0.0559$.}
\label{scalingMdeig}
\end{figure}

To check the consistency of the results with respect to the existence of
a transition within the RP$^2$ phase we can perform a Finite Size Scaling
study fitting $\Delta_\lambda L^{\beta/\nu}$ as a function of 
$(T-T_0)L^{1/\nu}$. Only $T_0=T_\mathrm{c}$ yields a reasonable fit (see
Fig.~\ref{scalingMdeig}).
Notice that
for $T$ close to $T_\mathrm{c}$ the definition of $\Delta_\lambda$ 
ceases to be
meaningful, as a large eigenvalue exists no more since the RP$^2$
magnetization fades away, and no good fit can be expected. However, for
most $T$ values (more precisely, for $T<0.05$) the fit is excellent, though
the $\eta$ and $\nu$ values are admittedly rather unusual
($\eta=-0.5$, $\nu=1.8$). The conclusion should be that there is no 
evidence for an O(2) breaking transition at any finite distance from
$T_\mathrm{c}$. A collapse of that transition over the RP$^2$-PM transition 
might occur.

\begin{figure}
\includegraphics[angle=90,width=\columnwidth]{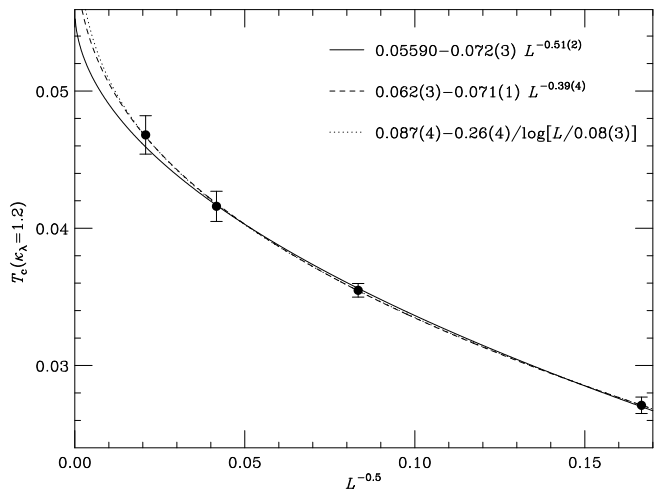}
\caption{Displacement of the critical temperature, defined as the point
where  $\kappa_\lambda$ takes a fixed value, as a function of size.
Fits suggest that there is no O(2)
breaking transition, at a temperature less than the RP$^2$-PM transition
temperature.}
\label{Binderfijolambda}
\end{figure}

A more quantitative analysis can be made studying the displacement of
the temperature at a fixed value of  $\kappa_\lambda$. In
Fig.~\ref{Binderfijolambda} we plot the obtained measures together with 
fits to several functional forms: a power law with $T\mathrm{c}$ fixed to the
value of Eq.~(\ref{tcrp2}), a three parameter power law, and a 
Kosterlitz-Thouless-like divergence. The results point again to no 
breaking of the O(2) symmetry inside the RP$^2$ phase.

\subsection{Interplay between Ferromagnetism, Antiferromagnetism, 
Temperature and an applied Magnetic Field in the low-doped La$_{1-x}$Sr$_x$MnO$_3$. }\label{HSUBSECT}

\begin{figure}
\includegraphics[angle=90,width=\columnwidth]{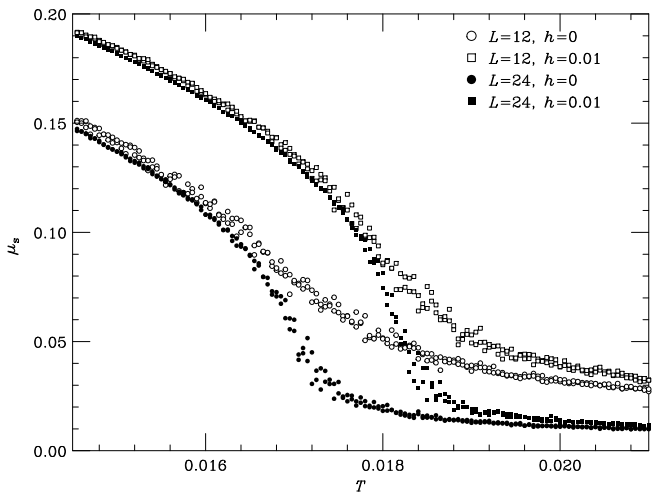}
\caption{Hysteresis along t$_2$. The transition occurs in the region where 
$\muVs$ changes suddenly, and the figure shows a movement of
$T_\mathrm{c}$ to higher values when $h=0.01$ is switched on 
($T_\mathrm{c}(h=0)=0.0171$). }
\label{fmfih}
\end{figure}
In a series of papers\cite{Uhlen,Nojiri,Wagner,Shun}
the interplay between FM, AFM, temperature and an applied magnetic field in
the low-doped La$_{1-x}$Sr$_x$MnO$_3$, mainly at $x$ close to 1/8, has been 
studied. We 
would like to point out some properties of our FM-FI phase transition (point 
t$_2$ in this paper) which
might help to understand phenomena which, in those references, are related to 
the FM-CO (Charge Ordered) phase transition, not fully understood so far.

 Roughly speaking, some of the mentioned phenomena are:
\begin{enumerate}

\item Resistance increases as $T$ decreases below $T_\mathrm{CO}$
(t$_2$ in our model).  In our simplified model, this corresponds to
the fact that, when crossing the FM-FI transition, odd and even spins
cease to be aligned, which makes conductivity via DEM harder.
\item The Charge Ordered phase grows larger when an external magnetic
field is applied. In our case, we have run a simulation with non zero
magnetic field to see how the transition line moves.
\item In the CO phase, at fixed temperature, the magnetization increases with 
an external magnetic field, just as in a FM phase.
\end{enumerate}

Let us now describe the physics of the FM-FI transition. Near the FM-FI 
transition, in the FM phase the ordering is symmetric with respect to the 
odd-even exchange 
and the field fluctuates at random around the total magnetization vector,
the fluctuations being larger than in the FI phase, as shown by measurements of
specific heats and susceptibilities made in both phases. More precisely, the 
magnetization increases as the temperature goes down from the FM to the FI 
phase, which can be explained by a diminution of fluctuations. In fact, one 
would expect that the magnetization should be smaller in the FI phase, with
fixed odds and evens on an open cone around the odd direction, than in the 
FM phase, where the evens lie on a narrower cone, with a larger projection
on the odd direction. Yet, the large fluctuations
in the FM phase destroy the even contribution to the magnetization. The FI 
vacuum consists then on the odd, practically frozen, sublattice, and the 
even sublattice, with spins on an open, but less fluctuating cone.

Let us now look in the FM phase close to the FI transition, and switch
on a weak magnetic field in the Z direction. This will have the
general effect of collimating the spins. In more detail, odd spins
shall freeze closer to the Z direction, which shall cause the even
sublattice to freeze on the cone, with smaller
fluctuations. Paradoxically, the collimating effect of the magnetic
field in the Z direction is to stabilize the cone, effectively opening
it, giving rise to a more FI-like ordering, i.e., {\em the FI phase
shall invade the FM phase, and the critical temperature shall rise.}
This phenomenon (see point 2) is accompanied by an increase in the
magnetization at fixed temperature in the FI phase (see point 3).

In order to check the correctness of the description, we have simulated in the
neighborhood of the 
transition with $h=0.01$, which does not alter the system properties, and have
run a hysteresis cycle at $J=-0.43$ in $L=12,24$ between $T=0.01$ 
and $T=0.025$ (i.e. along
t$_2$). A good observable for the transition is $\muVs$. The results at the two
$L$ values show that the finite size effects are negligible in front of the
change in $T_\mathrm{c}$ with $h$.

Fig.~\ref{fmfih} shows the result, which confirms that the inclusion of a 
magnetic field rises $T_\mathrm{c}$, causing the invasion of the FI ordering 
into regions which at $h=0$ were FM.

\section{Conclusions}\label{CONCLUSIONSSECT}

We have studied a simple model for double exchange interactions which
retains a good number of interesting properties. It exhibits a complex
phase diagram with ferromagnetic, ferrimagnetic phases, with their
staggered counterparts, and a segment-ordered phase.

We obtain quantitatively all phases with approximate calculations
(Mean Field and $1/N$ expansions), which can be contrasted with {\em
exact\/} Monte Carlo calculations. With Monte Carlo simulations we
obtain, in addition to the precise positions of the transitions,
information about their order. Our conclusion is that all transitions
seem to be second order, although an accurate determination of the
critical exponents is difficult and it is beyond the scope of this
paper.

We have studied in detail the {\em exotic\/} RP$^2$ phase
(segment-ordered), concluding that it is a single phase up to the
resolution allowed by the lattice sizes used in the simulation.  The
presence of a RP$^2$ phase up to $T=0$ is interesting from the
experimental point of view, since it can be confused with a PM or
glassy phase and consequently with a Quantum Critical Point.  We have
shown that the structure factor ($V$, in fig.~\ref{Fourier}) remains
bounded in the full Brillouin-zone .  Therefore the RP$^2$ phase
cannot be detected in neutron scattering experiments as a long-range
ordering, although the phase transition will show up as a maximum
(more precisely, a cusp) of the specific heat. A short-range ordering
would of course always be present. Since the critical exponent
$\alpha$ is negative, the Harris criterion\cite{HARRIS} implies that
our results are robust against disorder effects.  The RP$^2$ phase is
characterized by a breakdown of the even-odd symmetry and a remaining
O(2) symmetry, despite the fact that the measured critical exponents
for the RP$^2$ transition point to a total breakdown of the initial
O(3) symmetry. We have also discussed the effects of a magnetic field
on the ferromagnetic-ferrimagnetic transition, and we have discussed
its interplay with electrical conductivity.

\section*{Acknowledgments}
It is a pleasure to thank F. Guinea for useful conversation at the beginning
of this work. We have maintained interesting discussions with A. Pelissetto
regarding the large $N$ approximation. 
This work has been partially supported through
research contracts FPA2001-1813, FPA2000-0956, BFM2001-0718,
BFM2003-8532, PB98-0842 (MCyT) and HPRN-CT-2002-00307
(EU). V. M.-M. is a Ram\'on y Cajal research fellow (MCyT) and
S. J. is a DGA fellow.  We have used the PentiumIV cluster RTN3 at the
Universidad de Zaragoza for the simulations.

\appendix

\section{Large $N$ approximation}\label{LARGEN}

We write the model as
\begin{equation}
{\cal H}=-N \sum_{<i,j>} W(1+\vphi_i\cdot \vphi_j) \; .
\end{equation}
The Boltzmann weight is $\exp(-\cal{H})$ and
\begin{equation}
W(x)= J x+\sqrt{x} \; .
\end{equation}

Using the expression of the Dirac deltas (one to fix the spin modulus,
$\vphi_i^2=1$, and another to write that $x=\vphi_i\cdot\vphi_j$) 
in terms of functional integrals, we can write the
partition function of the model in the following way:\cite{CarPel,ZJ}
\begin{equation}
{\cal Z}\propto \int \dd[\,\rho, \lambda,\mu, \vphi\,]\ \ee^{N A} \; ,
\end{equation}
with  $A$, that action,  as:
\begin{eqnarray}
A&=&\frac{\beta}{2} \sum_{<i,j>} \left( \lambda_{ij} + \lambda_{ij}\, 
\vphi_i\cdot\vphi_j- \lambda_{ij}\; \rho_{ij} 
+ 2 W(\rho_{ij})\right)\nonumber\\
&&-\frac{\beta}{2}\sum_i\mu_i\left(\vphi_i^2- 1\right)\;.
\end{eqnarray}

As we are interested (in this part of the calculation) in paramagnetic
or/and ferromagnetic phases, we separate the spin in two pieces: the
first one parallel to the  symmetry breaking direction,
$\phi^\parallel$ (one degree of freedom), and the orthogonal part ($N-1$
degrees of freedom), $\vphi^\perp$.  At this point, the spins
have no definite modulus, and we can perform the functional
integration over the orthogonal part of spins (a Gaussian integral)
\begin{equation}
\int \dd[\vphi^\perp]\ \ee^{-\frac{1}{2} 
\vphi^\perp\cdot \hat R \vphi^\perp}
\propto \exp\left(-\frac{N-1}{2}
\,\mathrm{Tr}\, \log \hat R  \right)\; ,
\end{equation}
where $R^{ab}_{ij}$ is the propagator ($a,b=1,...,N-1$ and 
$i$ lives in
the three-dimensional lattice) and it is given by
\begin{equation}
R^{ab}_{ij}=\delta^{ab}\beta 
\left( \beta \mu_i \delta_{ij}-\frac12\sum_\nu\lambda_{ij} \delta_{i_\nu j} \right) \; .
\end{equation}
The sum runs back and forth along the 3 lattice axes and $i_\nu$ is the 
neighbor of site $i$ in the direction defined by $\nu$.
The trace, $\mathrm{Tr}$, is over the space and spin components. The quantity
$\frac12(\vphi^\perp\cdot \hat R \vphi^\perp)$ is the contribution to $A$ 
involving the orthogonal part of the spins (which is a quadratic
form with matrix $\hat R$).

In momentum space, $\hat R$ reads:
\begin{eqnarray}
R^{ab}(\bm q,\bm q')&=&\delta^{ab}\frac{\beta}{V}\sum_i 
\ee^{\ii (\bm q-\bm q')\cdot \bm r_i}\\\nonumber  
&&\times \bigg(  \mu_i 
-\frac12 \sum_{\nu}
[\lambda_{i i_\nu}
\ee^{\ii \bm q'\cdot\bm \nu} + \lambda_{i_{-\nu} i} \ee^{-\ii \bm q'\cdot 
\bm \nu}] \bigg),
\end{eqnarray}
where $\bm \nu=\bm r_{i_\nu}-\bm r_\nu$.
In the large $N$ technique we must maximize $A$.  In order to keep the
computation at its simplest level, we make an ansatz over the fields
$\lambda_{ij}$, $\mu_i$, $\rho_{ij}$, and $\phi^\parallel$: we
are assuming that we will describe under this ansatz translational
invariant phases, like paramagnetic and ferromagnetic ones. So, 
we will consider that all these fields are
independent of $x$ and $\mu$ and we will write them as $\lambda$,
$\rho$, $\mu$ and $\sigma$. Therefore, $A$ is
\begin{eqnarray}
\frac{A}{V}&=&\frac{\beta}{2} d \left[ \lambda (1-\rho)+\lambda \sigma^2
+2 W(\rho) \right]
+ \frac{\beta}{2} \mu (1-\sigma^2) \nonumber\\
&&-\frac{1}{2} \int \dd q \log \bigg[\mu - \lambda
\sum_\nu \cos q_\nu  \bigg]\;,
\end{eqnarray}
where $d$ is the dimension of space and  
\begin{equation}
\int \dd q \equiv \int_{[0,2 \pi)^d} \frac{\dd^d q}{(2 \pi)^d} =1\; .
\end{equation}
Hence, this computation is valid in paramagnetic/ferromagnetic
phases where we have translational invariance.  As usual we write
\begin{equation}
\hat{p}^2  \equiv 4 \sum_{\nu} \sin^2 (p_\nu/2) \; .
\end{equation}
The continuum limit of $\hat{p}^2$ is $p^2$, and so, we can define a
mass $m_0$:
\begin{equation}
m_0^2=\frac{2 \mu}{\lambda}-2 d \; .
\end{equation}
and  $A$ can be written as
\begin{eqnarray}
\frac{A}{V}&=&\frac{\beta}{2} d \left[ \lambda (1-\rho)+\lambda\sigma^2+2 W(\rho) \right]
+ \frac{\beta}{2} \mu(1-\sigma^2)\nonumber\\
&& -\frac{1}{2} \log \lambda 
-\frac{1}{2}\int \dd q \log \left[m_0^2+\hat{p}^2 \right] \; .
\end{eqnarray}
The saddle point equations are:
\begin{equation}
\beta d (1-\rho) +\frac{1}{\lambda} \left[ (m_0^2+2 d) I(m_0^2)-1 \right]
+d \beta \sigma^2 =0\;,
\end{equation}
\begin{equation}
\beta (1-\sigma^2)=\frac{2}{\lambda} I(m_0^2)\;,
\end{equation}
\begin{equation}
2 W'(\rho)=\lambda\;,
\end{equation}
\begin{equation}
\sigma(d \lambda-\mu)=0\;,
\end{equation}
where
\begin{equation}
I(m_0^2)\equiv\int \dd q \frac{1}{m_0^2+\hat{p}^2}\;.
\end{equation}

One solution is $\sigma=0$, the paramagnetic phase. We can find a second
order phase transition by fixing the mass $m_0$ to zero:
\begin{equation}
T I(0) =J +\frac{1}{2 \sqrt{\rho_0}}\;,
\end{equation}
where 
\begin{equation}
\rho_0=2-\frac{1}{2d I(0)}\;.
\end{equation}
In three dimensions $I(0)\simeq 0.2527$. So we have found the critical
line between the paramagnetic and ferromagnetic phases. This solution
is only valid in $J\ge -1/2$. It is easy to check that the $J=-1/2$
vertical line corresponds to infinite mass. So, with these formulas we cannot 
reach the region to the left of $J=-1/2$. Below we will see how to
solve this drawback.

We can try to connect this calculation with the $T=0$ results.  The
solution $\sigma \neq 0$ implies that $d\lambda=\mu$ and so
$m_0^2=0$. Notice that in a magnetized phase, $m_0$ has no longer the
meaning of a mass (hence, in this case, $m_0=0$ is not a signature of
criticality). In this case the complete solution is:
\begin{equation}
\rho^*=2-\frac{1-\sigma^2}{2 I(0) d}\;,
\end{equation}
and 
\begin{equation}
T=\frac{1-\sigma^2}{I(0)} \left[J +\frac{1}{2 \sqrt{\rho^*}} \right]\;.
\end{equation}
This last equation tells us what is the
magnetization $1-\sigma^2$ in a given point $(T,J)$ .  
In the interval $J> -1/2$ we obtain the
solution $\sigma=1$. In addition in the interval $J \in (-1/2,
-1/(2\sqrt{2}))$ a second solution with $\sigma<1$ appears. This is the
signature of the ferrimagnetic phase. Hence, we have recovered part of the 
 previous $T=0$ results.

As mentioned, above the previous calculation is valid only in 
paramagnetic/ferromagnetic phases. In order to manage the
paramagnetic/antiferromagnetic phases we use the following trick: we
change the sign of the odd spins and we leave
unchanged the even spins, so, the Hamiltonian reads:
\begin{equation}
{\cal H}=-N \beta \sum_{<i,j>} W(1-\vphi_x\cdot \vphi_y) \; .
\end{equation}
and following the technique outlined above, we obtain
the equations of the saddle point:
\begin{equation}
\lambda=-2 W^\prime(\rho)\;,
\end{equation}
\begin{equation}
1-\sigma^2= \frac{2 T}{\lambda} I(m_0^2)\;,
\end{equation}
\begin{equation}
d \rho= d(1-\sigma^2)-\frac{T}{\lambda} \left((2 d+m_0^2) I(0)-1\right)\;,
\end{equation}
\begin{equation}
\sigma(d \lambda-\mu)=0 \;.
\end{equation}
Again $m_0^2=2 \mu/\lambda-2 d$.

In the paramagnetic phase $\sigma=0$ is the solution and the equation
of the critical line is (obtained by fixing $m_0^2=0$):
\begin{equation}
-T I(0) =J +\frac{1}{2 \sqrt{\rho_0}}\;,
\end{equation}
where 
\begin{equation}
\rho_0=\frac{1}{2 d I(0)} \; .
\end{equation}

In addition $0<\sigma<1 $ is also a solution and so $d \lambda=\mu$
and this implies, as in the PM/FM computation, that $m_0^2=0$ The phase being a
(staggered) magnetized one, this does not imply criticality. The
solution is then:
\begin{equation}
\rho^*=\frac{1-\sigma^2}{2 I(0) d}\;,
\end{equation}
and 
\begin{equation}
T=-\frac{1-\sigma^2}{I(0)} \left[J +\frac{1}{2 \sqrt{\rho^*}} \right]\;.
\end{equation}
As in the PM-FM case, this last equation tells us what is the magnetization 
$1-\sigma^2$ in a given point $(T,J)$.
Again, in this part of the calculation we cannot reach the region
$J>-1/2$. The line $J=-1/2$ has again $m_0^2=0$.

Finally, we report the transition lines in terms of the temperature
measured in the Monte-Carlo simulation. Taking into account that
$T_{\mathrm{MC}}=T/N$, where $T$ is the temperature of the large $N$
calculation and fixing $N$ to 3, we obtain the FM-PM line:
\begin{equation}
T_{\mathrm{MC}}=1.2578\,J+0.5578\;,
\end{equation}
and the AFM-PM line
\begin{equation}
T_{\mathrm{MC}}=-1.2578\,J-0.793\;.
\end{equation}

\section{Mean Field Fourth order analysis}\label{FOURTH}

\begin{figure}
\includegraphics[angle=0,width=\columnwidth]{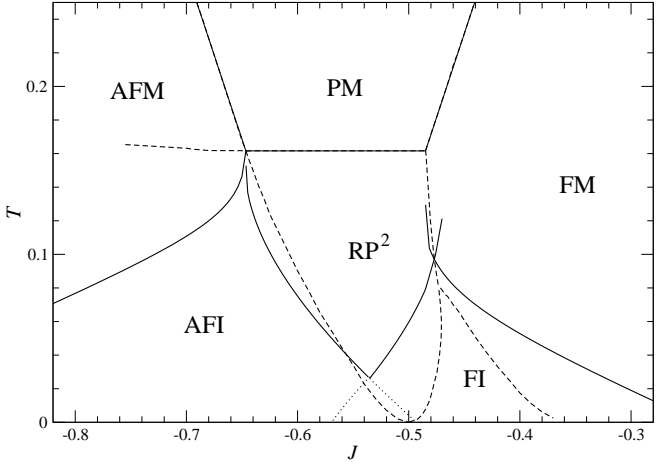}
\caption{Mean-Field phase diagram as obtained from the numerical minimization of the free-energy (dashed lines) and from the minimization of the free-energy
calculated to fourth order (full lines).  The dotted lines are artifacts of the
fourth-order approximation.
}
\label{fourthfig}
\end{figure}

We have extended our Mean Field power expansion analysis to fourth order,
so that we can find transitions where the paramagnetic phase is not involved.  
The analytical minimization with respect to all fields is a very hard
task. But we can face the problem restricting the parameter region,
using the essential fields that can describe the transition. First of
all, let us explore the transitions inside the ferromagnetic region
found in the second order analysis.

\begin{equation}
\Phi_{h \lambda_\st}(\lambda_\st)=\Phi(h^\mathrm{min},0,0,\lambda_\st)\,,
\end{equation}
where $h^{min}$ is the value of h where $\Phi_{h}(h)=\Phi(h,0,0,0)$ 
reaches the minimum.
We can expand
\begin{widetext}
\begin{equation}
\Phi_{h \lambda_\st}(\lambda_\st)=\Phi(h^\mathrm{min},0,0,0)+
a_{h \lambda_\st}(T,J)\lambda_\st^2+
b_{h\lambda_\st}(T,J)\lambda_\st^4+O(\lambda_\st^4)\;.
\end{equation}
Then, if $b_{h \lambda_\st}(T,J)$ is positive there is a stable minimum with 
non zero $\lambda_\st$ when
 $a_{h \lambda_\st}(T,J)$ is negative. Therefore, we find a transition
line when  $a_{h \lambda_\st}(T,J)=0$. In this case  $b_{h
  \lambda_\st}(T,J)>0$ if $T<0.31$ and the transition line between the
ferromagnetic and a ferrimagnetic phase, where $M$ and $M_\st$ are non
zero, is 
\begin{equation}
T_\mathrm{FM-FI}=-\frac{4 \left(20+83\sqrt2J+140J^2\right)}
{386\sqrt2+875 J+\sqrt{369392+971810\sqrt2J+1265425J^2}}\,.
\end{equation}
We can do a similar analysis inside the RP$^2$ phase. In this case we study
\begin{equation}
\Phi_{\lambda_\st h}(h)=\Phi(h,0,0,\lambda_\st^\mathrm{min})\,,
\end{equation}
and
\begin{equation}
\Phi_{\lambda_\st h_\st}(h_\st)=\Phi(0,h_\st,0,\lambda_\st^\mathrm{min})\,,
\end{equation}
obtaining the following transition lines
\begin{equation}
T_{\mathrm{RP}^2\mathrm{-FI}}=\frac{32\left(327+406\sqrt2J\right)}
{35\left(480\sqrt2+539J+\sqrt{-575136-768768\sqrt2J+290521J^2}\right)}\,,
\end{equation}

\begin{equation}
T_{\mathrm{RP}^2\mathrm{-AFI}}=\frac{32\left(283+406\sqrt2J\right)}
{35\left(-128\sqrt2+539J-\sqrt{929312+1148224\sqrt2J+290521J^2}\right)}\,.
\end{equation}

Finally, inside the antiferromagnetic phase we find a transition to an
antiferrimagnetic ordering minimizing  

\begin{equation}
\Phi_{h_\st \lambda_\st}(\lambda_\st)=
\Phi(0,h_\st^\mathrm{min},0,\lambda_\st)\,.
\end{equation}

The transition line is

\begin{equation}
T_\mathrm{FM-FI}=-\frac{4\left(404+795\sqrt2J+700J^2\right)}
{5\left(296\sqrt2+875J+\sqrt{463688+1085630\sqrt2J+1265425J^2}\right)}\,.
\end{equation}

The fourth-order phase-diagram is depicted in Fig.~\ref{fourthfig},
together with the numerical calculation of
Section.~\ref{MEANFIELDSECT}. Letting aside the FM-RP$^2$ line (which
is first-order in Mean-Field approximation), the results of the
fourth-order approximation are qualitative satisfying.

\end{widetext}


\begin{thebibliography}{99}

\bibitem{CVM99} J. M. D. Coey, M. Viret and S. von Molnar, Adv. in
Phys.  {\bf 48}, 167 (1999).

\bibitem{DAGOTTO} See e.g. E. Dagotto, T. Hotta and A. Moreo,
Phys. Rep. {\bf 344}, 1 (2001).

\bibitem{DAGOTTOBOOK}E. Dagotto, {\em Nanoscale Phase
Separation and Colossal Magneto resistence,} Springer Verlag (2002).

\bibitem{RAIZLETTER} J.M.~Carmona et al., Phys. Lett. B {\bf 560}, 140 (2003).

\bibitem{AMIT} See e.g., D. Amit, {\em Field Theory, the
Renormalization Group and Critical Phenomena\/}, World Scientific, 
Singapore (1984).

\bibitem{DEM}
C. Zener, Phys. Rev. {\bf 82}, 403 (1951); P. W. Anderson and
H. Hasegawa, {\em ibid.} {\bf 100}, 675 (1955); P. G. de Gennes, {\em
ibid.} {\bf 118}, 141 (1960).

\bibitem{Verges01}
J. A. Verges, V. Mart\'{\i}n-Mayor and L. Brey,
Phys. Rev. Lett.  {\bf 88}, 136401 (2002).

\bibitem{Alonso01} J. L. Alonso, J. A. Capit\'an, L. A. Fern\'andez,
F. Guinea, V. Mart\'{\i}n-Mayor, Phys. Rev. B {\bf 64}, 54408 (2001).

\bibitem{Alonso00a}
J. L. Alonso, L. A. Fern\'andez, F. Guinea, V. Laliena, V. Mart\'{\i}n-Mayor,
Phys. Rev. B {\bf 63},  64416 (2001).

\bibitem{Alonso00b} J. L. Alonso, L. A. Fern\'andez, F. Guinea,
V. Laliena, V. Mart\'{\i}n-Mayor, Phys. Rev. B {\bf 63}, 54411 (2001).

\bibitem{Alonso00c}
J. L. Alonso, L. A. Fern\'andez, F. Guinea, V. Laliena,
V. Mart\'{\i}n-Mayor, Nucl. Phys. B {\bf 596},  587  (2001).

\bibitem{KuMa} S. Kumar and P. Majumdar, cond-mat/0305345.

\bibitem{TSAILANDAU} 
Shan-Ho Tsai, D.P. Landau, J. of Mag. Mat. Mag. {\bf 226-230}, 650 (2001)
650 and J. of App.  Phys. {\bf 87}, 5807 (2000).

\bibitem{QCP} S. Sachdev, {\em Quantum Phase Transitions}, Cambridge
University Press, Cambridge (1999); Science {\bf 288}, 475 (2000).

\bibitem{Burgy01} J. Burgy, M. Mayr, V. Mart\'{\i}n-Mayor, A. Moreo
and E. Dagotto, Phys. Rev. Lett. {\bf 87}, 277202 (2001).

\bibitem{Alonso02}
J. L. Alonso, L. A. Fern\'andez, F. Guinea, V. Laliena and 
V. Mart\'{\i}n-Mayor, Phys. Rev. B {\bf 66}, 104430 (2002).

\bibitem{DAGOTTO03}
See E. Dagotto, preprint cond-mat/0302550 and Ref.~\onlinecite{DAGOTTOBOOK}.

\bibitem{RP2D3}H. G. Ballesteros, L. A. Fern\'andez, V. Mart\'{\i}n-Mayor, and
         A. Mu\~noz Sudupe, Phys. Lett. B \textbf{78}, 207 (1996); 
	 Nucl. Phys. B \textbf{483}, 707 (1997).  

\bibitem{ROMANO} S. Romano, Int. J. of Mod. Phys. B {\bf 8}, 3389  (1994).

\bibitem{SHROCK} G. Korhing and R.E. Shrock, Nucl. Phys. B {\bf 295}, 36
(1988).

\bibitem{Uhlen} S. Uhlenbruck et al.,
Phys. Rev. Lett. {\bf 82}, 185 (1999). 

\bibitem{Nojiri} H. Nojiri, K. Kaneko, M. Motokawa, K. Hirota, Y. Endoh and 
K. Takahashi,
Phys. Rev. B {\bf 60}, 4142 (1999). 

\bibitem{Wagner} P. Wagner, I. Gordon, S. Mangin, V.V. Moshchalkov, 
Y. Bruynseraede, L. Pinsard and A. Revcolevschi,
Phys. Rev. B {\bf 61}, 529 (2000). 

\bibitem{Shun} Shun-Qing Shen, R.Y. Gu, Qiang-Hua Wang, Z.D. Wang and X.C. Xie,
Phys. Rev. B {\bf 62}, 5829 (2000). 

\bibitem{AZARIA} P. Azaria, B. Delamotte and T. Jolicoeur,
Phys. Rev. Lett. {\bf 64}, 3175 (1990); 
P. Azaria, B. Delamotte, F. Delduc and T. Jolicoeur,
Nucl. Phys. {\bf B408}, 485 (1993).

\bibitem{KAWAMURA} See for example H. Kawamura, Can. J. Phys. 
{\bf 79}, 1447 (2001).

\bibitem{NOCHIRAL} M. Tissier, B. Delamotte and D. Mouhanna,
Phys. Rev. Lett. {\bf 84},  5208 (2000); M. Itakura,
J.Phys. Soc. Jpn. {\bf 72}, 74 (2003).

\bibitem{ChiralN23}
A. Pelissetto, P. Rossi and E. Vicari, Phys. Rev. {\bf B65},  020403
(2002) and references therein.

\bibitem{COOPER} 
  B.Cooper, B. Freedman and D. Preston, Nucl. Phys. B {\bf 210}, 210 (1992). 
  J.K. Kim, Phys. Rev. Lett. {\bf 70}, 1735 (1993). 
  S. Caracciolo, R.G. Edwardas, A. Pelissetto and A.D. Sokal, 
  Nucl. Phys. B {\bf 403}, 475 (1993).

\bibitem{parisi}See eg. G. Parisi, {\it Statistical Field Theory\/}, 
Addison Wesley, New York (1988).

\bibitem{DeGennes}
P.-G.De Gennes,
Phys.Rev. Lett. {\bf 118}, 1 (1960).

\bibitem{SDHV} H. G. Ballesteros and V. Mart\'{\i}n-Mayor,
Phys. Rev. E {\bf 58}, 6787 (1998).

\bibitem{cocientes2}
H. G. Ballesteros, L.A. Fern\'andez, V. Mart\'{\i}n-Mayor, 
A. Mu\~noz Sudupe, G. Parisi and J. J. Ruiz-Lorenzo,
J. Phys. A: Math. Gen. {\bf 32}, 1 (1999).

\bibitem{FERRSWEN}A. M. Ferrenberg and R. H. Swendsen,
  Phys. Rev. Lett. {\bf 61}, 2635 (1988).

\bibitem{ISPERC}
H. G. Ballesteros, L. A. Fern\'andez, V. Mart\'{\i}n-Mayor, 
A. Mu\~noz Sudupe, G. Parisi and J. J. Ruiz-Lorenzo,
J. Phys. A: Math. Gen. {\bf 32}, 1 (1999).

\bibitem{XYCLASS}
M. Campostrini, M. Hasenbusch, A. Pelissetto, P. Rossi and E, Vicari,
Phys. Rev. B {\bf 63}, 214503 (2001).

\bibitem{HASENBUSCH}
M. Hasenbusch, J. Phys. A {\bf 34}, 821 (2001).

\bibitem{LOMBARDO}
L.A. Fern\'andez, M.P. Lombardo,
J.J. Ruiz-Lorenzo and A. Taranc\'on, Phys. Lett. {\bf B277},  485 (1992).

\bibitem{HARRIS}
A. B. Harris, J. Phys. {\bf C7},  1671 (1974).

\bibitem{CarPel}
S. Caracciolo and A. Pelissetto, Phys. Rev. {\bf E66}, 016120 (2002).

\bibitem{ZJ}
J. Zinn-Justin, {\em Quantum Field Theory and Critical Phenomena\/}, 
Oxford Science Publications, Oxford (1990).


\end{thebibliography}
\end{document}